\documentclass{aa}   
\usepackage{enumerate, graphicx}
\usepackage{float}
\usepackage{txfonts}
\usepackage{pdflscape}
\usepackage{multirow}

\usepackage{hyperref}
\hypersetup{
    colorlinks=true,
    linkcolor=black,
    urlcolor=black,
    citecolor=blue,
    }
\usepackage{comment}
\usepackage{lscape}
\usepackage{cancel}
\usepackage[shortlabels]{enumitem}
\usepackage{siunitx}
\usepackage{tabularx}
\usepackage{lineno}

\DeclareSIUnit \h {\mbox{$h$}}
\DeclareSIUnit \parsec {pc}
\DeclareSIUnit \Msun {\mbox{M$_{\odot}$}}
\DeclareSIUnit \Lsun {\mbox{L$_{\odot}$}}
\DeclareSIUnit \century{century}
\DeclareSIUnit \year{yr}

\newcommand{\jwst}{\textit{JWST}}

\usepackage{natbib}
\bibpunct{(}{)}{;}{a}{}{,} 

\begin{document}

\title{Faint galaxies in the Zone of Avoidance revealed by \jwst/NIRCam}  

\author{J. L. Nilo-Castell\'on\inst{1}  \and  M. V. Alonso\inst{2,3} \and L. Baravalle\inst{2,3} \and  C. Villalon\inst{2,3} \and C. N. A. Willmer\inst{10} \and C. Valotto\inst{2,3} \and M. Soto\inst{4}  \and  D. Minniti\inst{5,6}  \and M. A. Sgró\inst{11} \and I. V. Daza-Perilla\inst{7,8}  \and H. Cuevas Larenas\inst{1} \and A. Ramirez\inst{1} \and J. Alonso-Garc\'{i}a\inst{9} \and P. Marchant Cortés\inst{1} \and F. Milla Castro\inst{1} } 

\institute{  
Departamento Astronom\'ia, Facultad de Ciencias, Universidad de La Serena. Av. Raul Bitran 1305, La Serena, Chile.
\and
Instituto de Astronom\'{\i}a Te\'orica y Experimental, (IATE-CONICET), Laprida 854, X5000BGR, C\'ordoba, Argentina.
\and
Observatorio Astron\'omico de C\'ordoba, Universidad Nacional de C\'ordoba, Laprida 854, X5000BGR, C\'ordoba, Argentina.
\and
Instituto de Investigación en Astronomía y Ciencias Planetarias, Universidad de Atacama, Av. Copayapu 485, Copiapó, Chile.
\and
 Instituto de Astrof\'isica, Facultad de Ciencias Exactas, Universidad Andr\'es Bello, Av. Fernandez Concha 700, Las Condes, Santiago, Chile.
 \and
Vatican Observatory, V00120 Vatican City State, Italy.
\and
Center for Space Science and Technology, University of Maryland, Baltimore County, 1000 Hilltop Circle, Baltimore MD 21250, USA.
\and
Center for Research and Exploration in Space Science and Technology, NASA/Goddard Space Flight Center, Greenbelt, MD 20771, USA.
\and
Centro de Astronom\'ia (CITEVA), Universidad de Antofagasta, Av. Angamos 601, Antofagasta, Chile
\and
Steward Observatory, University of Arizona 933N Cherry Avenue Tucson AZ 85751 USA.
\and
Instituto de Altos Estudios Espaciales “Mario Gulich” (CONAE – UNC), Argentina
}

   \date{Received xx; accepted xx}

\abstract
{The zone of avoidance (ZoA) remains one of the last frontiers in constructing a comprehensive three-dimensional map of the Universe. Galactic extinction, stellar crowding, and confusion noise have historically limited the detection of background galaxies in these regions, with implications for large-scale structure and cosmological measurements.}
{We assess the capability of the James Webb Space Telescope (\jwst) Near Infrared Camera (NIRCam) to detect extragalactic sources in a heavily contaminated region of the Milky Way.}
{We analyzed \jwst/NIRCam wide-filter images of NGC~3324 with a customized implementation of SExtractor v2.28. Sources were detected in the F444W band, cross-matched with F090W and F200W, and validated against recent DAOPHOT point spread function (PSF) photometry. A refined sample was obtained through full width at half maximum (FWHM)–signal-to-noise ratio (SNR) criteria and visual inspection.}
{We identified 102 galaxies across the \textit{JWST}/NIRCam field of view. The magnitude (F444W) distribution is bimodal, with $\sim10$\% brighter than $m_{\mathrm{F444W}}<15$~mag and $\sim60$\% in the range $17<m_{\mathrm{F444W}}<19$~mag. Typical sizes are ${\rm FWHM}\approx6.5^{\prime\prime}$, from compact to extended systems with isophotal areas up to $\sim2000$~pixels ($\sim7.9~\mathrm{arcsec}^2$). Morphologies span from compact to spiral and lenticular systems, including a compact group at the eastern edge of the field. We also report the detection of \textit{transnebular galaxies}, visible through the most opaque regions of the molecular cloud.}
{These results demonstrate the potential of \jwst/NIRCam to probe extragalactic sources through highly obscured Galactic regions, opening new avenues for mapping large-scale structures across the ZoA.}


   \maketitle
\section{Introduction}\label{introduction}
The reconstruction of the three-dimensional distribution of matter in the Universe requires uniform and complete coverage across the sky, since only then can observational maps be meaningfully compared with theoretical expectations. Within modern cosmology, these expectations are set by the $\Lambda$ Cold Dark Matter ($\Lambda$CDM) paradigm \citep[e.g.][]{White1978,Blumenthal1984,Springel2005}, where small primordial density fluctuations grow hierarchically through gravitational instability of cold dark matter, giving rise to the observed cosmic web of filaments, clusters, and voids \citep[e.g.][]{Zeldovich1970,Bond1996,VandeWeygaert2008,OKane2024,TojeiroKraljic2025}. Yet, despite this well-established paradigm, reconstructions remain observationally incomplete, limiting our ability to confront theoretical predictions with data.

In response to this challenge, successive generations of wide-area redshift surveys have sought to recover the morphology of the cosmic web by providing three-dimensional maps of the nearby and intermediate-redshift sky. Landmark contributions include the Two-degree Field Galaxy Redshift Survey (2dFGRS; \citealt{Colless2001}; see also \citealt{vandeWeygaert2009}), the Sloan Digital Sky Survey (SDSS; \citealt{York2000}; see also \citealt{Tegmark2004}), and the 6dF Galaxy Survey (6dFGS; \citealt{Jones2009}), while deeper or more recent programs such as GAMA \citep{Driver2011}, VIPERS \citep{Guzzo2014}, eBOSS \citep{Alam2021}, and DESI \citep{DESICollab2023} have extended this mapping to higher redshifts and larger volumes, reinforcing the empirical connection between observed galaxy distributions and the growth of large-scale structure.

One region, however, has persistently resisted such efforts: the Zone of Avoidance (ZoA; \citealt{Proctor1878}; see also \citealt{KraanKorteweg2000}). It arises from the combination of high concentrations of Galactic dust—responsible for the observed interstellar extinction \citep{Schlegel1998,Schlafly2011}—and severe stellar crowding along the plane. Although it covers only $\sim$10 per cent of the sky in optical surveys, the ZoA obscures structures that may be essential to a complete description of the cosmic web, with consequences that manifest both in the nearby Universe and at higher redshifts.

In the local volume, the ZoA coincides with the projected positions of major mass concentrations, including the Norma \citep{Woudt2008} and Vela \citep{KraanKorteweg2017} superclusters, whose continuity across the plane remains only partially established. Such incompleteness propagates into uncertainties in reconstructions of the nearby gravitational field \citep{Lahav1994}, in the characterization of bulk flows \citep{Courtois2013}, and in the derivation of cosmological parameters from local structures \citep[e.g. estimates of $H_0$ and $\Omega_m$;][]{Tully2016, Riess2016, Planck2020, Pantheon2022, Clocchiatti2024}. 

Beyond the local volume, the contiguous mask imposed by the ZoA spans angular scales comparable to those of the transverse baryon acoustic oscillation (BAO) feature ($\sim 3^{\circ}$ at $z \sim 0.85$ in the Dark Energy Survey Year 6; \citealt{DESCollab2024}). Theoretical and methodological studies have long established that such masking couples Fourier modes through the survey window and amplifies the integral-constraint effect \citep{Peebles1973, Peacock1991, Hamilton1993, Wilson2017, DeMattia2019}. Empirical analyses show that incomplete sky coverage introduces measurable biases in two-point statistics and BAO determinations \citep{Beutler2011, Beutler2014, Avila2025}, underscoring that robust measurements of large-scale clustering demand a consistent treatment of the Zone of Avoidance. Alternative formulations to mitigate window-function and integral-constraint effects are discussed in \citet{Feldman1994}, \citet{Landy1993}, and \citet{Eisenstein2007}.

Over the last two decades, multi-wavelength strategies have sought to mitigate these limitations. Optical surveys led by Ren{\'e}e Kraan-Korteweg explored low-extinction windows but became ineffective beyond $A_B > 3$ mag \citep{KraanKorteweg2000}. Radio surveys, particularly those targeting the 21-cm H\,{\sc i} line, provided an alternative means of probing the obscured sky. The Parkes H\,{\sc i} ZoA Survey \citep{Donley2005, Henning2010, StaveleySmith2016}, a blind program with the 64-m Parkes telescope \citep{StaveleySmith1996}, detected 883 galaxies with recessional velocities up to 12,000 km\,s$^{-1}$. These data traced extensions of the Great Attractor \citep{Woudt2007}, Puppis \citep{KraanKorteweg1999}, and the Local Void \citep{Tully2022}, while also uncovering new concentrations (e.g. NW1–3, CW1–2) consistent with wall-like structures in the southern sky. In the northern ZoA, \citet{KraanKorteweg2018} used the Nançay Radio Telescope \citep{Lequeux2010} to detect 220 new galaxies, reveling new filaments, reinforced the delineation of known structures in Monoceros and Puppis, and traced the continuation of the Perseus–Pisces Supercluster across the galactic plane \citep{Ramatsoku2016}. Despite these advances, the reach of optical and radio campaigns remained constrained by survey depth and wavelength sensitivity.

Near-infrared surveys overcame many of these limitations and represented a decisive step in mitigating incompleteness in the Zone of Avoidance (ZoA). The Two Micron All Sky Survey (2MASS; \citealt{2MASS}) provided uniform near-infrared coverage, revealing thousands of previously obscured galaxies and reducing the effective ZoA for luminous systems to only a few percent of the sky \citep{Jarrett2000}. The subsequent 2MASS Redshift Survey (2MRS; \citealt{Huchra2005}) obtained redshifts for nearly 45,000 galaxies with $K_{s}\leq11.75$ mag, reaching an initial completeness of 97.6\% and later approaching full coverage through targeted follow-up of ZoA galaxies \citep{Macri2019}. These data enabled the first contiguous three-dimensional mapping of the local Universe across the Galactic plane, forming the basis for density- and velocity-field reconstructions. Analyses of 2MRS confirmed the alignment of the galaxy dipole with the CMB dipole \citep{Maller2003}; at $\sim50\,h^{-1}$ Mpc, the 2MRS flux-weighted dipole lies within $\sim12^{\circ}$ of the CMB dipole direction, diverging to $\sim21^{\circ}$ by $\sim130\,h^{-1}$ Mpc \citep{Erdogdu2005}. Furthermore, the catalog enabled the identification of over 13,000 groups across 91\% of the sky and the reconstruction of the three-dimensional density field between 3000 and 10,000 km s$^{-1}$ \citep{Tully2015}.

The advent of the VISTA Variables in the Vía Láctea surveys (VVV; \citealt{vvv}) and its extension VVVX \citep{VVVX} provided the first homogeneous, sub-arcsecond, multi-epoch near-infrared coverage of the low-latitude bulge and disc. VVV mapped $\sim562\,\mathrm{deg}^{2}$ and VVVX expanded the footprint to $\sim1\,700\,\mathrm{deg}^{2}$, with median seeing of $0.8^{\prime\prime}$ and a $5\sigma$ point-source depth of $K_{s}\simeq16.9$~mag. This expanded coverage enabled systematic searches for extragalactic sources at low Galactic latitudes. Early analyses combined automated extraction with {\sc SExtractor}+{\sc PSFEx} and visual inspection, yielding trousands of previously unreported galaxy candidates \citep{Baravalle2018}. These efforts culminated in the VVV Near-Infrared Galaxy Catalogue (NIRGC~I), which reported 5\,563 visually confirmed galaxies, $\sim99\%$ of them new \citep{Baravalle2021}. Subsequent releases adopted machine-learning pipelines: NIRGC~II employed a six-channel convolutional neural network together with an XGBoost classifier to identify 1\,003 galaxies in the northern disc \citep{Daza2023}, while NIRGC~III extended the search to the southern disc, yielding 167\,559 candidates down to $K_{s,0}=16$~mag over $1\,080\,\mathrm{deg}^{2}$ \citep{Alonso2025}. Within this framework, the first galaxy cluster discovered by the survey, VVV-J144321-611754, was spectroscopically confirmed in  \citet{Baravalle2019}, and complementary analyses based on Voronoi tessellations and minimum-spanning trees identified additional  candidate groups and clusters \citep{Soto2022}, reinforcing the evidence for large-scale structures that cross the Galactic plane. Systematic searches were also extended into the bulge, the most heavily extincted and crowded sector of the ZoA, where \citet{Galdeano2021}, \citet{Duplancic2024}, and \citet{Galdeano2025} identified galaxy overdensities and filamentary structures, including the Ophiuchus cluster \citep{Galdeano2022}.

Despite these advances, mapping galaxies behind the Milky Way remains incomplete and strongly modulated by the Galaxy's baryonic component, as indicated by residual extinction patterns and detection asymmetries (e.g. Fig.~3 of \citealt{Macri2019}; Fig.~13 of \citealt{Duplancic2024}; Fig.~10 of \citealt{Alonso2025}).

Several studies have attempted to infer the continuity of large-scale structures across the Galactic plane through indirect or constrained reconstructions. Constrained cosmological realizations and statistical galaxy fields predict plausible morphologies for the hidden cosmic web behind the ZoA \citep[e.g.][]{Sorce2017,McAlpine2022,Pfeifer2023}. These analyses indicate that major superclusters and filaments may extend across the plane, yet the lack of direct detections leaves their topology and mass distribution weakly constrained. This highlights the need of observational strategies capable of penetrating the most obscured sectors of the ZoA, a regime now accessible only through space-based facilities.

Here we report the identification of 102 galaxies located behind a section of the heavily obscured star-forming complex NGC~3324, using deep near-infrared imaging obtained with \jwst/NIRCam. These detections demonstrate that high-resolution, space-based infrared observations can effectively overcome the extinction barrier of the Galactic disc, enabling direct access to the extragalactic background through regions traditionally excluded from large-scale structure analyses. The paper is structured as follows: Section~\ref{background} describes the NGC~3324 star-forming region and the \jwst \ imaging; Section~\ref{data} presents the observational framework and datasets; Section~\ref{selection} details the morpho-photometric segmentation and source selection; Section~\ref{result} reports the resulting sample of background galaxies; and Section~\ref{discusion} discusses the main results and their implications for studies of the Zone of Avoidance.

\begin{figure}[ht]
\centering
\includegraphics[width=0.45\textwidth]{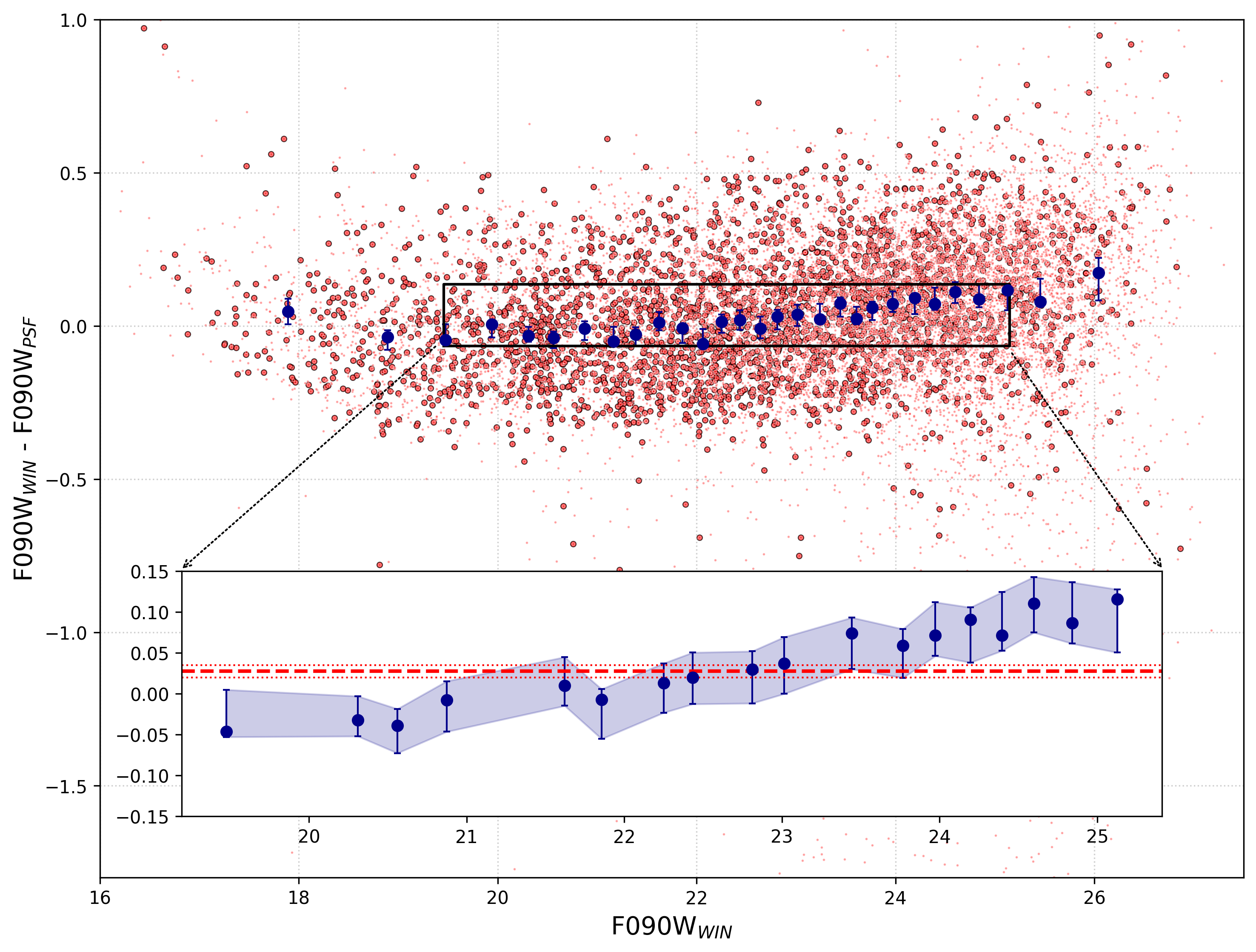}
\includegraphics[width=0.45\textwidth]{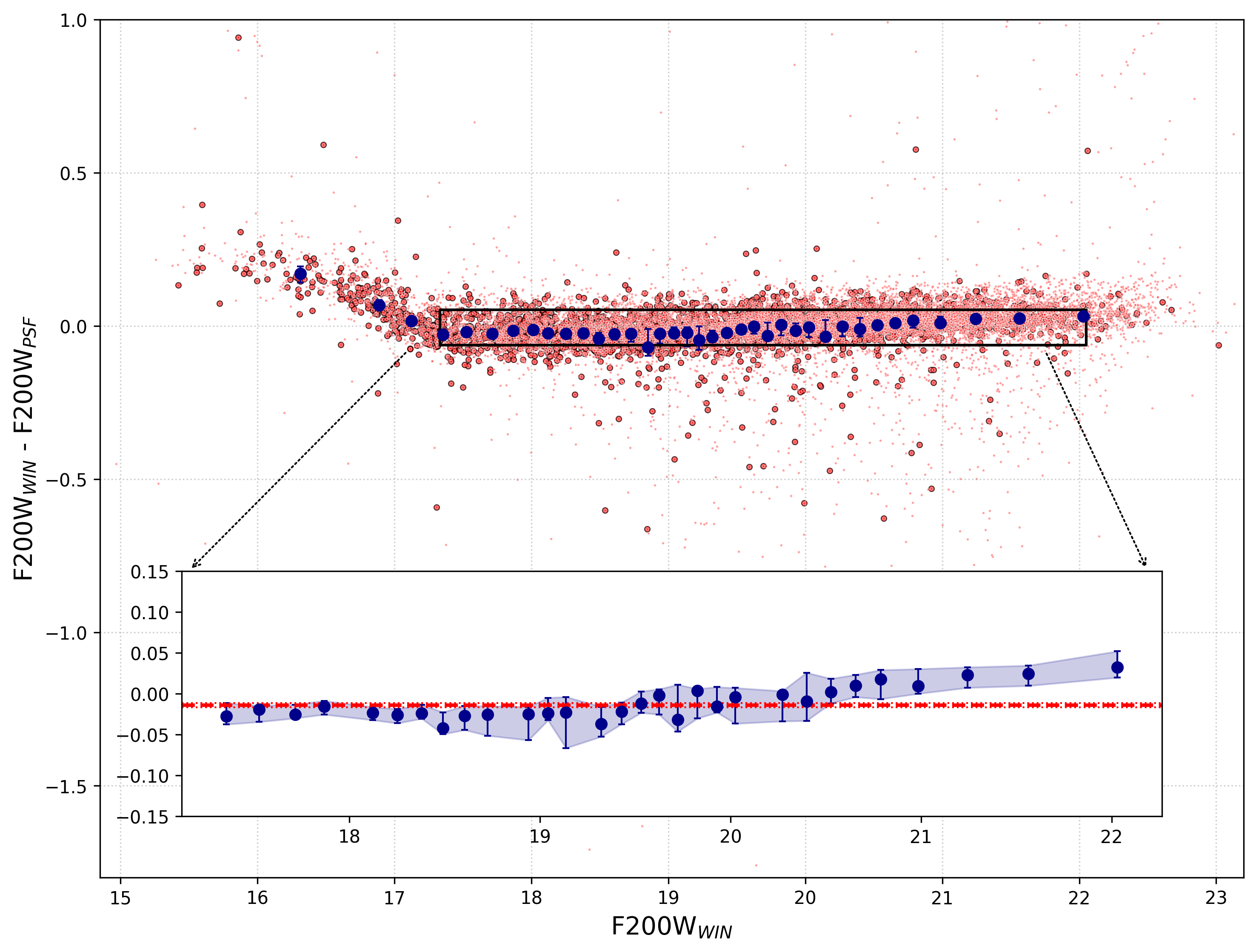}
\includegraphics[width=0.45\textwidth]{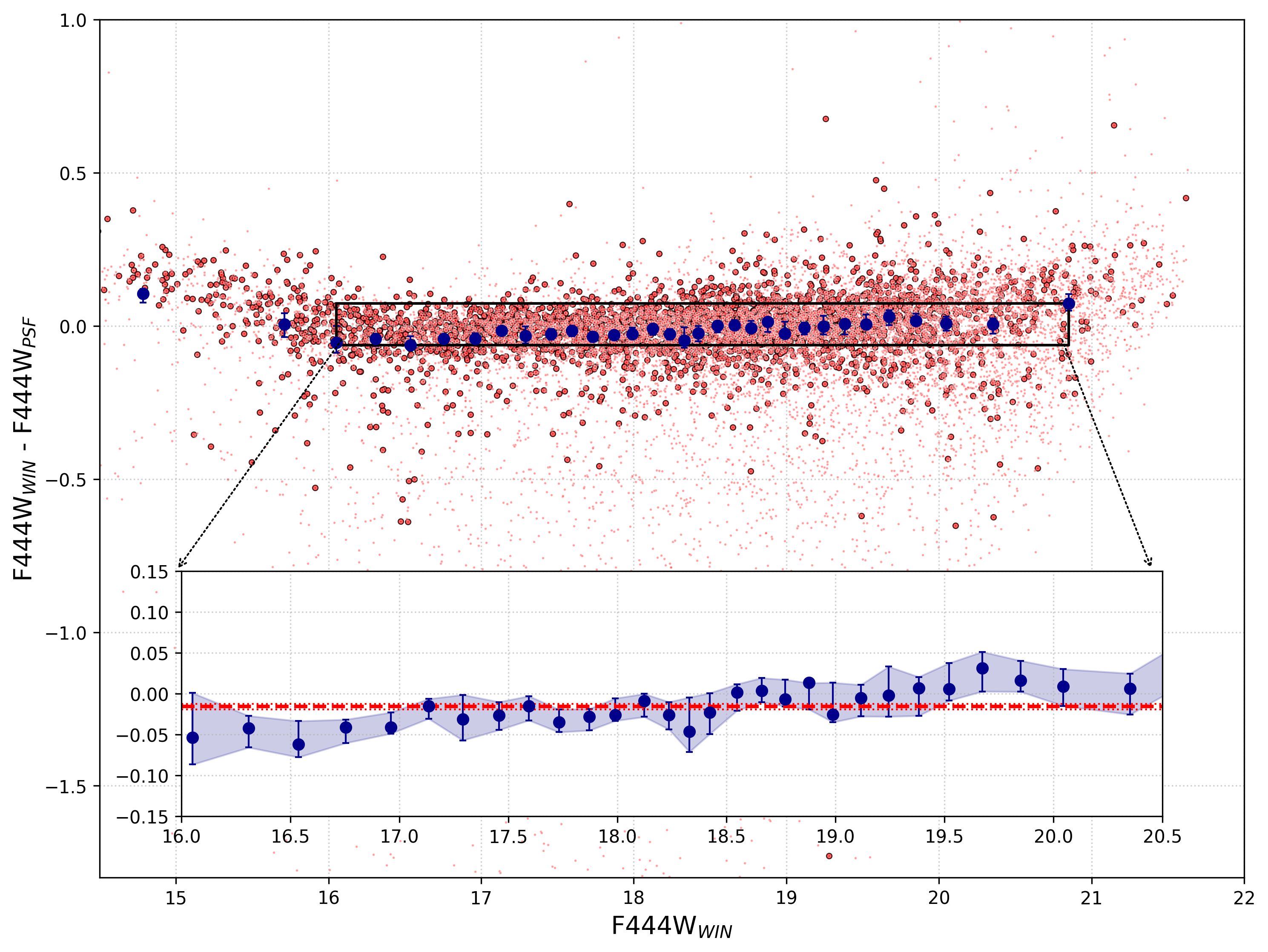}
\caption{Photometric calibration for the F090W (top), F200W (middle), and F444W (bottom) bands. Matched sources are shown as faint red dots. The subset of point-like sources (eccentricity $< 0.01$) used for calibration is indicated by red circles. Blue circles with error bars represent bin-averaged values computed via the Binscatter method. The lower inset zooms into the region used for regression, delimited by dashed lines. Horizontal dashed lines indicate the linear fits used to compute the final magnitude offsets.}
\label{fig:fotometria}
\end{figure}

\section{The Star-Forming Region NGC~3324 and the \jwst Images}\label{background}

NGC~3324 is an H\,\textsc{ii} region at the northwestern boundary of the Carina Nebula complex, with an angular extent of $\sim11\arcmin$ and a heliocentric distance of 2.35~kpc \citep[][and references therein]{Goppl2022}. Although historically treated as part of the broader Carina system, morphological discontinuities and filamentary connections toward Carina suggest a partially detached molecular structure linked by bridges of dust and gas \citep{Preibisch2012,Roccatagliata2013,Reiter2023}.

The western interface, the ``Cosmic Cliffs,'' delineates the illuminated inner wall of a cavity carved by radiative and mechanical feedback from massive stars. The ionizing field is dominated by the O-type systems HD~92206 and CPD$-57^\circ$~3580, which inject momentum into the surrounding medium and shape the local photoionization front \citep{Smith2007}.

A portion of NGC~3324 was imaged by the {\it James Webb Space Telescope} ({\it \jwst}) with the Near-Infrared Camera (NIRCam; \citealt{Beichman2012,Rieke2023}) and the Mid-Infrared Instrument (MIRI; \citealt{Rieke2015,Wright2015}) as part of program PID~2731 \citep{Pontoppidan2022}. The region’s radiative stratification, high surface brightness, and morphological richness were leveraged to demonstrate the performance of {\it \jwst} in angular resolution, dynamic range, and spectral coverage.

Beyond their demonstrative role, these observations delivered significant scientific results: \citet{Reiter2022} identified 31 protostellar outflows (including seven Herbig--Haro objects) previously inaccessible at shorter wavelengths; \citet{Dewangan2023} resolved intertwined filamentary substructures down to $\sim$4500\,au; and \citet{Crompvoets2024} constructed a deep catalog of $\sim$19{,}500 sources and isolated $\sim$450 candidate young stellar objects using probabilistic classification. Collectively, these findings establish NGC~3324 as a benchmark environment for dissecting star formation under strong radiative and mechanical feedback.

\begin{figure}[ht]
\centering
\includegraphics[width=0.45\textwidth]{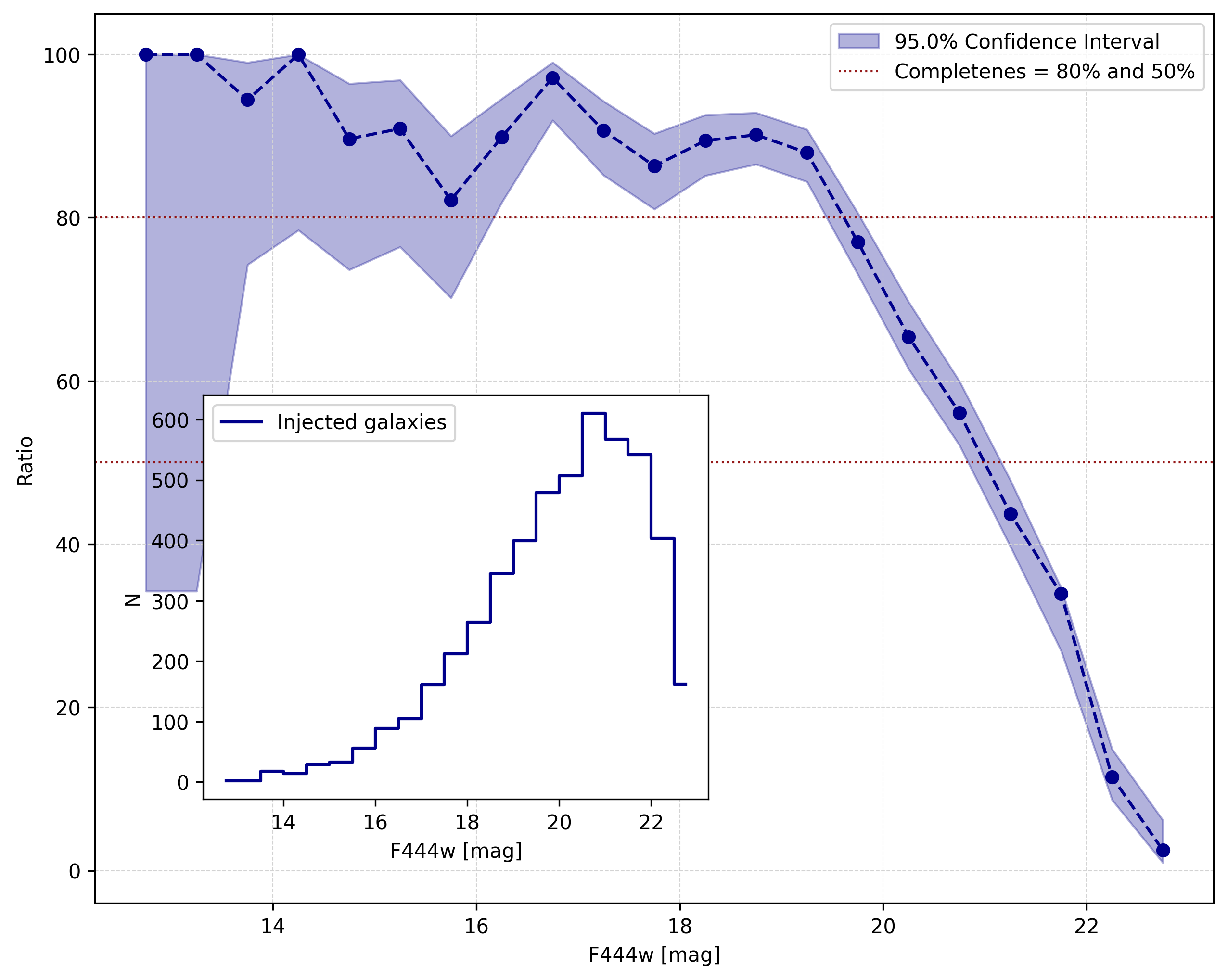}
\caption{Completeness as a function of apparent magnitude in the {\tt F444W} band. The blue curve shows the recovery fraction of 5000 injected synthetic galaxies, with shaded regions indicating the 95\% confidence interval derived from binomial statistics. Horizontal dashed lines mark the 80\% and 50\% completeness thresholds. The inset panel shows the magnitude distribution of the injected sources, sampled from a beta distribution to mimic the shape of the observed luminosity function.}
\label{fig:completitud}
\end{figure}

\section{Observational Framework and Imaging Datasets}\label{data}

This work is part of a comprehensive effort to detect and characterize galaxies obscured by the Zone of Avoidance (ZoA), utilizing deep near-infrared imaging from the VISTA Variables in the Vía Láctea \citep[VVV;][]{vvv} and its extended campaign VVVX \citep{VVVX}. This initiative produced the VVV Near-Infrared Galaxy Catalog (VVV NIRGC), comprising three main components: the inner southern Galactic disk (VVV NIRGC I; \citealt{Baravalle2021}), sectors of the northern disk (VVV NIRGC II; \citealt{Daza2023}), and the southern disk (VVV NIRGC III; \citealt{Alonso2025}). Collectively, these catalogs contain 174,116 galaxies, the vast majority of which have been identified for the first time.

\subsection{NIRCam Imaging and Filter Rationale}\label{catalogs}

Following the methodological protocol defined in the aforementioned VVV and VVVX analyses, galaxy candidates are selected from source catalogs generated using {\tt SExtractor} v2.28 \citep{Bertin+Arnouts1996}. The selection applies a set of morpho-photometric criteria optimized to distinguish extended extragalactic sources from the foreground stellar population.

We analyze broad-band {\it \jwst}/NIRCam observations obtained with the {\tt F090W}, {\tt F200W}, and {\tt F444W} filters. While previous analyses of these images have focused on Galactic feedback and star formation processes \citep[e.g.,][]{Dewangan2023}, the same data reach depths that enable the detection of large numbers of faint, compact sources beyond the sensitivity of ground-based surveys. A subset of these sources is consistent with background galaxies projected behind the complex. \citet{Crompvoets2024} report a catalog of $\sim$19,500 detections in this field, with limiting depths of 28.6 mag ({\tt F090W}), 26.2 mag ({\tt F200W}), and 22.8 mag ({\tt F444W}, AB, 5$\sigma$).

The choice of broad-band filters is motivated by their higher integrated throughput compared to narrow-band alternatives, which directly increases the probability of detecting photons from faint extragalactic sources. This follows from their significantly larger effective bandwidths ($\Delta \lambda \sim 0.2$--1.0 $\mu$m for {\tt F090W}--{\tt F444W} versus $\Delta \lambda \sim 0.02$--0.04 $\mu$m for narrow-band filters), resulting in deeper sensitivities for continuum sources at fixed exposure time \citep[see NIRCam Instrument Handbook;][]{JDox}.

\begin{figure}
\centering
\includegraphics[width=0.45\textwidth]{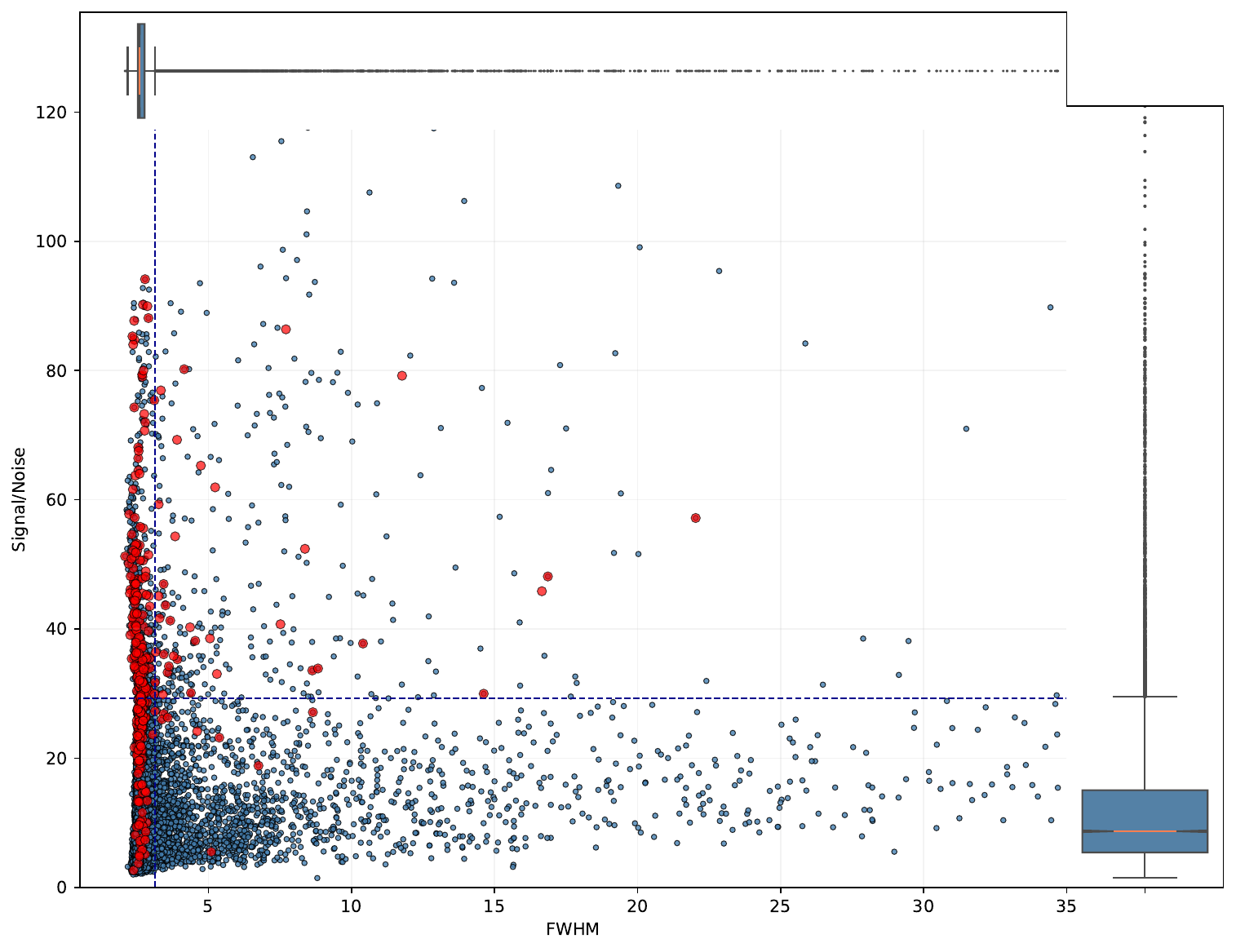}
\caption{The parameter space used to distinguish point-like from extended sources is shown. Blue dots represent all detections in the field, while red circles indicate sources matched with \textit{Gaia} stars. Vertical and horizontal lines mark the third quartile of the distributions in \texttt{FWHM} and \texttt{SNR\_WIN}, respectively, providing empirical thresholds for source separation. 
 }
\label{fig:separacion}
\end{figure}

\subsection{Source Extraction and Catalog Construction}\label{catalogs}

Source extraction was performed using {\tt SExtractor} ({\tt SE}) on the {\tt F444W} image, chosen as the detection frame because of its greater depth and reduced stellar contamination relative to shorter wavelengths. Independent extractions in the {\tt F090W} and {\tt F200W} bands provided complementary photometry, and catalogs were cross-matched with a tolerance of $0.1\arcsec$ based on \texttt{ALPHAWIN\_J2000} and \texttt{DELTAWIN\_J2000} coordinates, consistent with the internal astrometric accuracy of the mosaics.  

The principal challenge in applying {\tt SE} to the NGC~3324 field was the treatment of a spatially non-uniform background. Inaccurate modeling induces false positives near ionization fronts, diffraction spikes from bright stars, and filamentary emission. To mitigate these effects, several non-default parameters were adopted: (i) \texttt{BACK\_PEARSON} = 3.5 to suppress interpolation across correlated gradients common in PAH-rich regions, (ii) \texttt{BACKPHOTO\_THICK} = 12 pixels to improve local background estimation, and (iii) \texttt{BACK\_FILTTHRESH} = 3 to restrict filtering to uncorrelated fluctuations. A mesh size of \texttt{BACK\_SIZE} = 32 pixels and a filtering kernel of \texttt{BACK\_FILTERSIZE} = 3 were selected to balance sensitivity and background continuity. Spurious detections were further excluded by rejecting sources with \texttt{DETECT\_MAXAREA} $>2000$ pixels (corresponding to areas larger than expected for genuine galaxies at this depth) or exceeding the saturation threshold (\texttt{SATUR\_LEVEL} $>20000$ ADU).  

Post-extraction filtering removed entries flagged as unreliable. Sources with negative fluxes or with critical detection flags ---saturation (\texttt{FLAGS\_WIN} = 4), truncated isophotes (\texttt{FLAGS\_WIN} = 8), or corrupted photometry (\texttt{FLAGS\_WIN} = 32)--- were discarded. The final catalog contains 18,029 sources with astrometric and morphometric parameters from {\tt F444W} and photometry in {\tt F090W}, {\tt F200W}, and {\tt F444W}. This number is consistent with the $\sim$19,500 sources reported by \citet{Crompvoets2024}, with the small difference attributable to more restrictive filtering of flagged detections.  

\begin{figure}[htbp]
\centering
\includegraphics[width=0.45\textwidth]{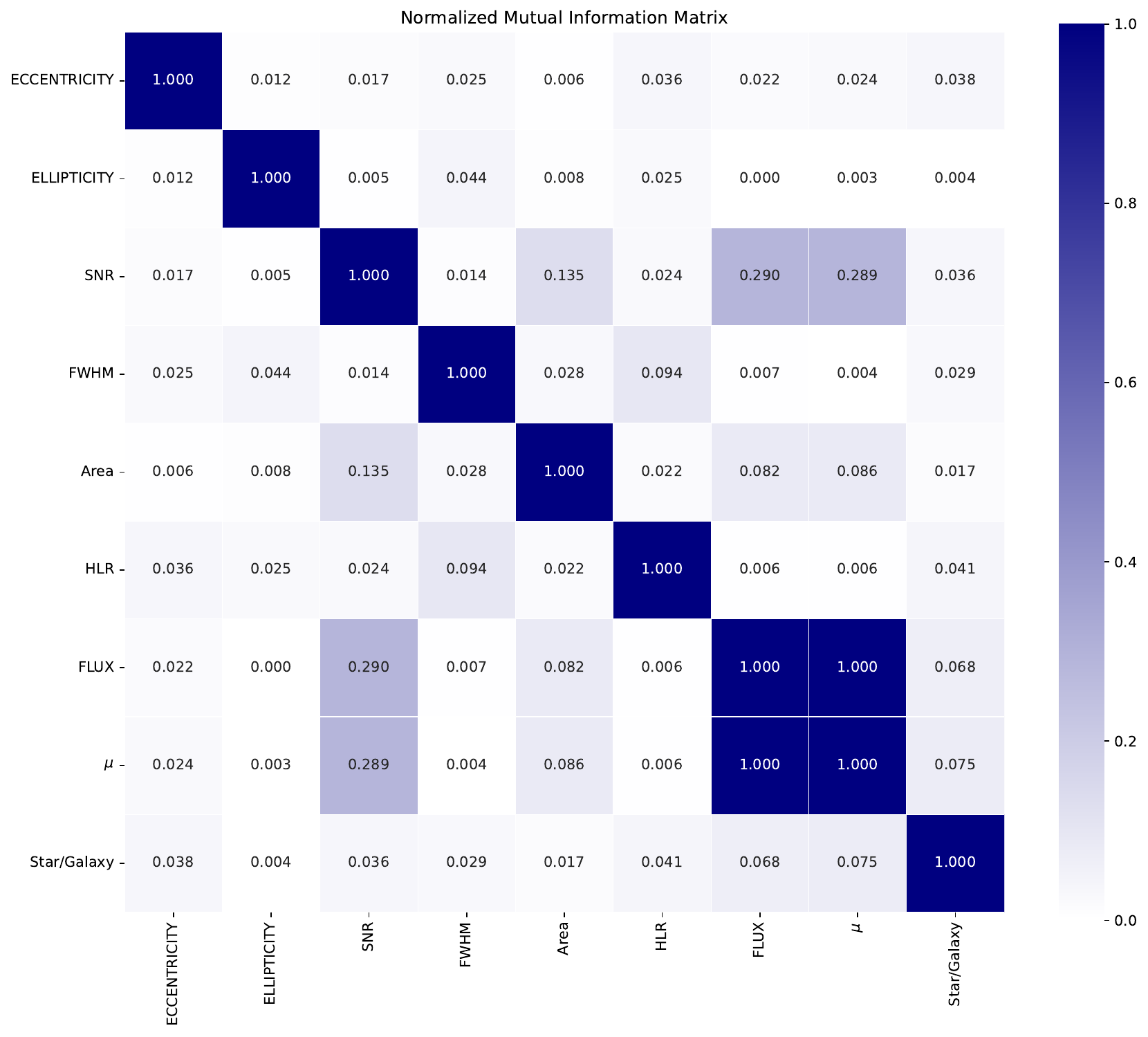}
\caption{Mutual information matrix among morphometric parameters. Color intensity reflects the degree of statistical dependence, from 0 (independent) to 1 (fully dependent).}
\label{fig:matrix}
\end{figure}

\subsection{Photometry Validation}\label{calibrations}

Photometric measurements were obtained from the pipeline \texttt{\_i2d} mosaics, which are provided in surface--brightness units (MJy\,sr$^{-1}$). To enable \texttt{SExtractor} photometry, which requires instrumental counts, we computed effective AB zeropoints that incorporate the pixel scale and the absolute calibration keywords in the image headers. This procedure converts the surface--brightness units into calibrated fluxes per pixel and ensures full consistency with the AB magnitude system. The resulting effective zeropoints are 26.29\,mag for {\tt F090W}, 25.58\,mag for {\tt F200W}, and 22.39\,mag for {\tt F444W}. These values are in agreement with those expected from the official \jwst/NIRCam calibration files for mosaicked products, after accounting for the adopted pixel scale.

Two complementary photometric estimators were extracted for every detected source and included in the scientific catalog: \texttt{MAG\_WIN} and \texttt{MAG\_AUTO}. \texttt{MAG\_WIN} is computed using Gaussian--weighted circular apertures scaled to each source’s full width at half maximum (FWHM). By design, these measurements capture of order half of the total flux while minimizing sensitivity to local background fluctuations, which makes them particularly valuable for point--spread--function (PSF) cross--checks and for validating photometric zeropoints. \texttt{MAG\_AUTO} corresponds to Kron--like elliptical apertures defined by the second--order moments of each source’s light distribution. This estimator is better suited to the irregular and non-circular morphologies characteristic of background galaxies: unlike fixed or circular apertures, Kron apertures provide a more stable approximation to the total flux of extended systems, while retaining robustness in the presence of crowding and structured backgrounds. Unless explicitly stated, all subsequent analyses in this work adopt \texttt{MAG\_AUTO} as the fiducial proxy for total flux.

To validate the accuracy of our photometric catalogues, we compared our \texttt{MAG\_WIN} measurements with the deep PSF photometry of \citet{Crompvoets2024}, obtained using \textsc{DAOPHOT}. Sources were cross--matched with a $0.2\arcsec$ tolerance, selected as a compromise between the intrinsic single--source astrometric precision of \jwst\ ($\sim$20--30 mas) and the larger centroiding uncertainties introduced by crowding and spatially structured nebular backgrounds in this field. This yielded 14{,}173 common detections. As part of this validation, we further examined calibration diagnostics by restricting the sample to sources with nearly pointlike morphology, defined by an eccentricity $<0.1$ computed from the eigenvalues of the second--moment covariance matrix (\texttt{CXXWIN\_IMAGE}, \texttt{CXYWIN\_IMAGE}, \texttt{CYYWIN\_IMAGE}). This criterion excludes elongated or distorted PSFs while retaining a representative set of stellar calibrators, which delineate a tight locus around $\Delta$[mag]$\approx 0$ across all filters, with increased scatter in {\tt F090W} consistent with the more complex short--wavelength backgrounds (see Fig.~\ref{fig:fotometria}).

Magnitude offsets were quantified using the binscatter approach \citep{Cattaneo2024} to average residuals in magnitude bins, followed by robust linear regression with RANSAC \citep{Fischler1981} to suppress outliers and isolate the central trend. This procedure yielded more stable results than standard sigma--clipping in the presence of non-Gaussian residuals. For {\tt F090W} we obtained mean/median offsets of $0.011$/$0.021$\,mag with a dispersion of $0.006$\,mag; for {\tt F200W}, $-0.010$/$-0.014$\,mag with $0.002$\,mag dispersion; and for {\tt F444W}, a mean/median of $0.072$/$0.071$\,mag. The small mean--median differences in {\tt F200W} and {\tt F444W} indicate symmetric residuals, while the larger offset at {\tt F090W} suggests mild skewness consistent with short--wavelength systematics. These offsets are within the expected absolute photometric accuracy envelope for \jwst\ and support the adopted calibration strategy.

\begin{figure}[htbp]
\centering
\includegraphics[width=0.45\textwidth]{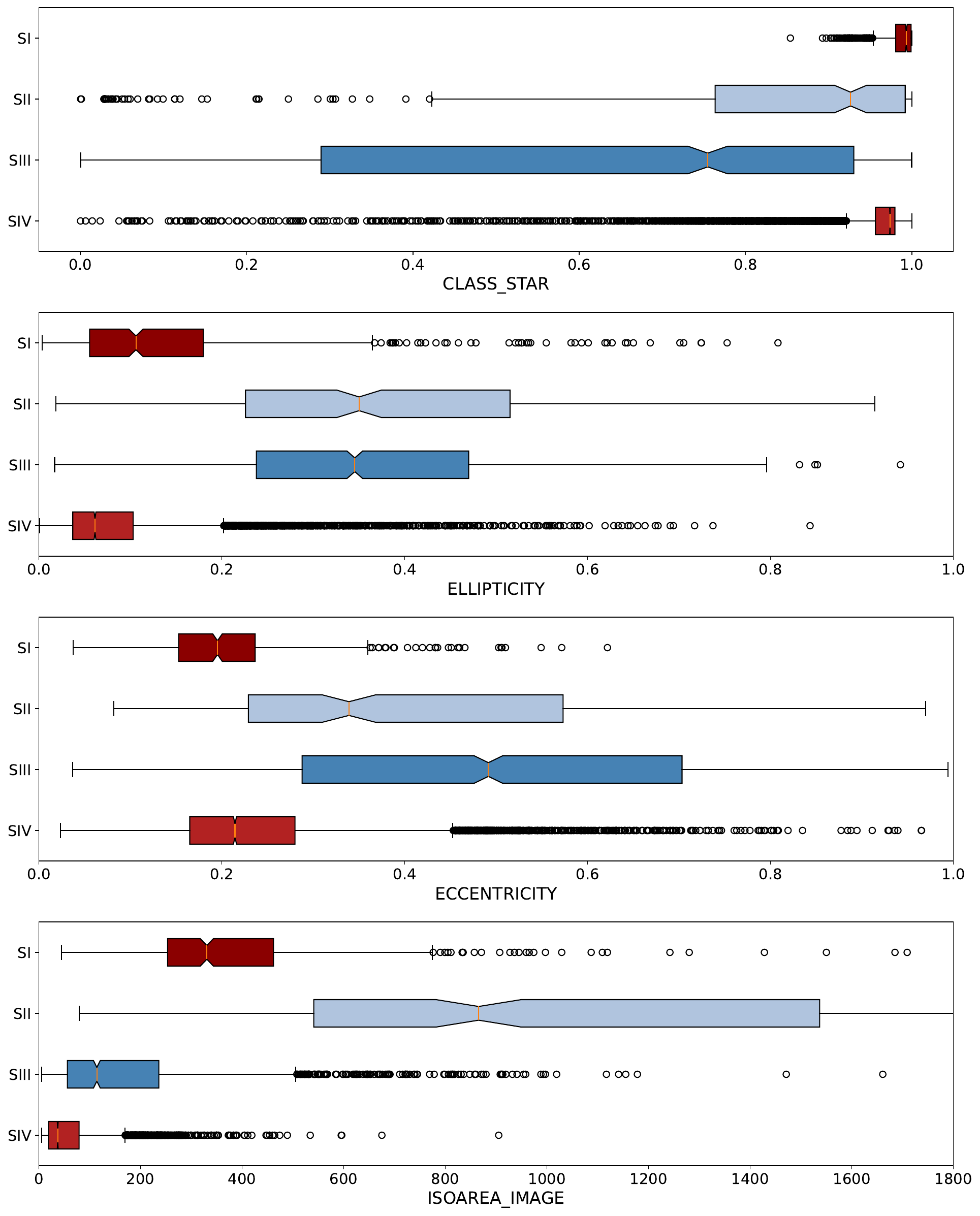}
\caption{Boxplot summary of \texttt{CLASS\_STAR}, \texttt{ELLIPTICITY}, \texttt{ECCENTRICITY}, \texttt{ISOAREA\_IMAGE} across the four sectors. Orange notches mark median values. Outliers are shown as open circles. Color codes correspond to sector classification as described in the text.}
\label{fig:boxplots}
\end{figure}

\subsection{Completeness and Photometric Depth}

Catalog completeness was quantified by means of an injection--recovery experiment on the {\tt F444W} image. A total of 5000 synthetic galaxies were randomly distributed across the full frame and subjected to the same extraction pipeline used in the primary analysis. 

Galaxy models were generated with the \texttt{Sersic2D} function from the \texttt{astropy.modeling.models} module, combining bulge+disc components representative of typical extragalactic systems. The bulge effective radii were drawn uniformly between 1 and 5 pixels, with ellipticities matched to the empirical mean of the field and random position angles in $[0,\pi]$. Disc scale lengths ranged from 5 to 20 pixels, inclinations from face-on to edge-on, and uniformly distributed position angles. Bulge-to-total flux ratios ($B/T$) varied continuously from 0 to 1, generating a morphologically diverse ensemble. All synthetic sources were convolved with a Gaussian kernel matching the empirical FWHM of the {\tt F444W} image, 
measured from isolated unsaturated stars as $ {\rm FWHM} = 0.20''$ (corresponding to $\sim3.2$~pixels at the NIRCam long-wavelength scale).  Modified frames were reprocessed with \texttt{SExtractor} using the same configuration as for the scientific catalog, and recovery fractions were computed in 0.5~mag bins.

Input magnitudes were sampled from a beta distribution with $\alpha = 8$ and $\beta = 2$, producing an asymmetric distribution over $13 < {\tt F444W} < 23$ mag with a prominent mode and extended faint tail \citep{merlin2022}. As shown in the inset of Figure~\ref{fig:completitud}, the reduced number of bright injected galaxies (${\tt F444W}<16$ mag) explains the larger binomial uncertainties in that regime.  

The completeness curve for {\tt F444W} is presented in Figure~\ref{fig:completitud}. The catalog reaches 80\% completeness at ${\tt F444W} = 19.5$ mag and 50\% at ${\tt F444W} = 21$ mag. Confidence intervals (95\%) were calculated with the Wilson score method. Extinction effects were not incorporated into the injection models, so the derived limits represent intrinsic detectability under uniform conditions. Nevertheless, the 50\% threshold coincides with the limiting sensitivity reported by \citet{Crompvoets2024} from PSF photometry on the same dataset, indicating that our results are coherent with previous analyses and capture the effective depth of the observations.  

\section{Morpho-photometric Segmentation and Source Selection}\label{selection}

A critical step in the construction of extragalactic catalogs across the Zone of Avoidance is the secure separation between point-like and extended sources. This classification directly governs the reliability of the sample. In NGC~3324 the problem is compounded because many extended detections correspond to local nebular substructures \citep{Reiter2022, Dewangan2023, Crompvoets2024}. The resolving power of \jwst/NIRCam further amplifies this degeneracy by revealing structure even in compact Galactic sources.

We distinguish point-like from extended sources in the joint plane of full width at half maximum (\texttt{FWHM}) and Gaussian–window signal-to-noise (\texttt{SNR\_WIN}). For a stable PSF, unresolved sources cluster at low \texttt{FWHM} and high \texttt{SNR\_WIN}, whereas extended sources occupy larger \texttt{FWHM} with typically lower \texttt{SNR\_WIN}.

Empirical validation was obtained by cross-matching {\tt F444W} detections with \textit{Gaia}~DR3 \citep{GaiaDR3}, yielding 1,363 associations. A high-confidence stellar subset of 309 sources was isolated with \texttt{RUWE}$<1.4$, $\varpi>0.5$~mas, and $0.5<{\tt BP-RP}<3$. About 75\% of this subset lies at \texttt{FWHM}$<4$, consistent with PSF-like loci.

Segmentation thresholds were defined from the third quartiles of the \texttt{FWHM} and \texttt{SNR\_WIN} distributions (Figure~\ref{fig:separacion}). Vertical and horizontal boundaries at $\texttt{FWHM}=3.124$ and $\texttt{SNR\_WIN}=29.257$ partition the plane into four sectors (\textbf{SI}--\textbf{SIV}). 

\begin{figure}[ht]
\centering
\includegraphics[width=0.5\textwidth]{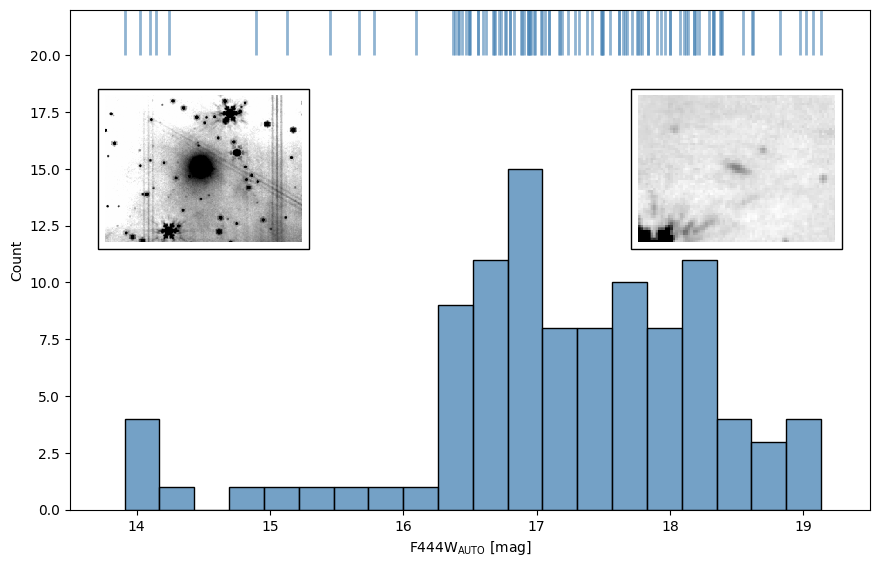}
\caption{Distribution of apparent magnitudes (\texttt{F444W\_AUTO}) for the galaxies detected in our study. Vertical ticks above the histogram indicate the magnitude of each galaxy. Two inset images display the brightest and faintest galaxies in the sample, providing visual references for the boundaries of the detected population. All magnitudes were measured using the Gaussian-weighted \texttt{WIN} aperture in the F444W band.}
\label{fig:magnitudes}
\end{figure}

To characterize each sector we examined \texttt{SExtractor} morphology. Because several descriptors are correlated projections of the same light distribution, we performed an information-theoretic independence analysis. Figure~\ref{fig:matrix} shows the normalized mutual-information (MI) matrix computed over all detections; values near zero indicate weak dependence. Evaluating all four-parameter subsets, the combination \texttt{ECCENTRICITY}, \texttt{ELLIPTICITY}, \texttt{ISOAREA\_IMAGE} (isophotal area at the detection threshold, expressed in both pixels and arcsec$^2$), and \texttt{CLASS\_STAR} minimizes the mean pairwise MI (0.014), significantly below the global median ($\sim$0.08), and is adopted for interpretation.

Figure~\ref{fig:boxplots} summarizes the distributions of these four descriptors per sector. \textbf{SI} and \textbf{SIV} are dominated by sources consistent with the instrumental PSF, i.e. unresolved at the angular resolution of NIRCam. Median \texttt{CLASS\_STAR} values are 0.993 and 0.974 with narrow interquartile ranges; median ellipticities remain below 0.11 and eccentricities below 0.22. Differences in median \texttt{ISOAREA\_IMAGE} (330 pixels in \textbf{SI} vs. 37 in \textbf{SIV}) reflect variations in angular size and surface brightness within compact populations, not a change of class.

By contrast, \textbf{SII} and \textbf{SIII} contain sources that are morphologically resolved relative to the NIRCam PSF, i.e. extended systems with broader and asymmetric parameter distributions. Median \texttt{CLASS\_STAR} values decrease to 0.926 (\textbf{SII}) and 0.754 (\textbf{SIII}); ellipticity medians rise to 0.350 and 0.345, and eccentricities to 0.339 and 0.491, indicating departures from circular symmetry. \textbf{SII} hosts the most extended, high-SNR systems (median \texttt{ISOAREA\_IMAGE} 865 pixels), while \textbf{SIII} comprises fainter, more compact extended sources (median 114 pixels), consistent with lower surface brightness.

Guided by this segmentation, we restrict subsequent analysis to the extended-source domains \textbf{SII} and \textbf{SIII}, and we apply ${\tt F444W\_WIN}<21$ (the 80\% completeness threshold) together with \texttt{CLASS\_STAR}$<0.5$ to limit stellar contamination. The resulting catalog contains 631 sources: 77 in \textbf{SII} and 554 in \textbf{SIII}. Each entry was visually inspected in $5\times5$-pixel cutouts from {\tt F090W}, {\tt F200W}, and {\tt F444W}; this quality-control step, applied after quantitative selection, confirmed extragalactic morphology in ambiguous cases.


\section{Background Galaxies behind NGC~3324: Results from \jwst/NIRCam}\label{result}

We identified 102 galaxies within the $7.38'\times4.23'$ ($\sim31~\mathrm{arcmin^2}$) \jwst/NIRCam mosaic covering the northwestern region of NGC~3324. The mosaic combines data from both NIRCam modules, with native pixel scales of $0.031''~\mathrm{pix^{-1}}$ in the short-wavelength channel and $0.063''~\mathrm{pix^{-1}}$ in the long-wavelength channel. All sources are detected in both the {\tt F444W} and {\tt F200W} bands, while 86 exhibit measurable flux in {\tt F090W}, illustrating the reduced impact of Galactic extinction at longer wavelengths. Astrometric and morphometric parameters were measured in the {\tt F444W} image, whereas photometry was extracted independently for each band.
A visual summary of all identified galaxies is presented in Appendix~\ref{fig:mosaic}.

\begin{figure*}[ht]
\centering
\includegraphics[width=1\textwidth]{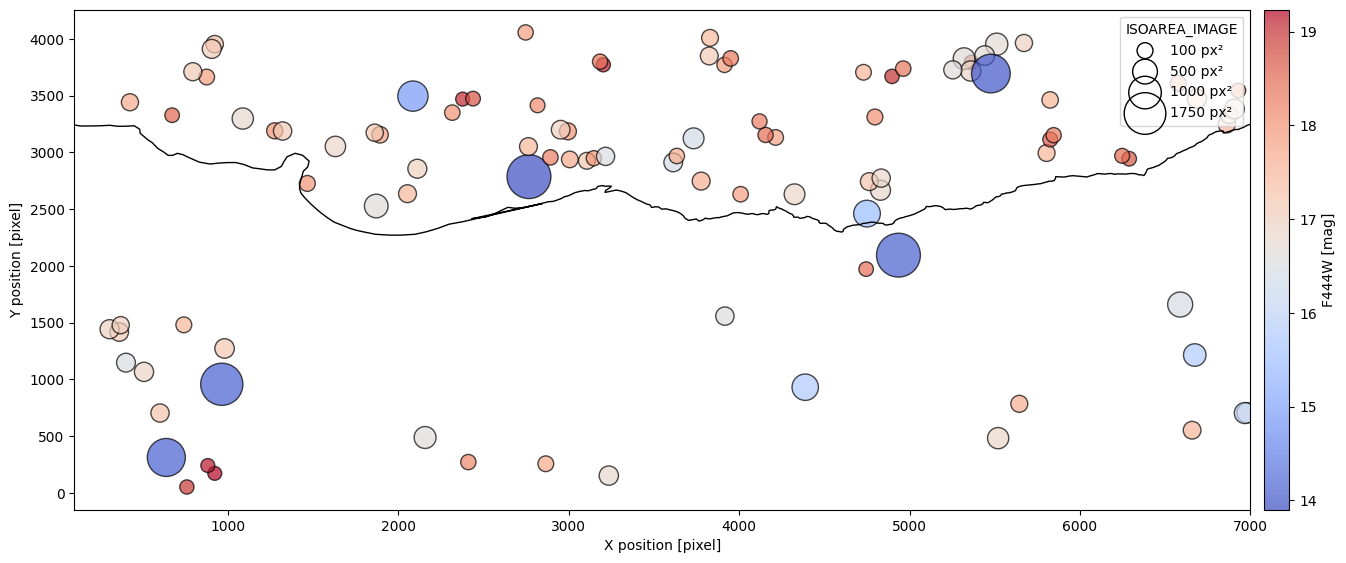}
\caption{Projected positions of the 102 galaxies detected in the \jwst/NIRCam observations of NGC~3324, shown in native detector coordinates. Marker color encodes {\tt F444W\_AUTO} magnitude (darker tones correspond to brighter sources), and marker size scales with the isophotal area ({\tt ISOAREA\_IMAGE}) in pixel units. The black line outlines the approximate boundary of the high-extinction region associated with the molecular cloud. A size reference for {\tt ISOAREA\_IMAGE} is included in the legend. The mosaic covers a field of view of $7.38'\times4.23'$ ($\sim31~\mathrm{arcmin^2}$). This figure provides a visual reference for the spatial distribution of the detected galaxies across the field.}
\label{fig:bubble}
\end{figure*}

Figure~\ref{fig:bubble} shows the locations of the 102 galaxies detected in the NGC~3324 field, displayed in detector $X$ and $Y$ coordinates. The figure is intended as a reference map illustrating the relative positions of the galaxies within the \jwst/NIRCam footprint, without further quantitative interpretation.

As mentioned, we adopted \texttt{MAG\_AUTO} as the reference magnitude, as it provides a consistent estimate of total flux for extended systems. The sample spans \texttt{F444W\_AUTO} magnitudes from 13.7 to 19.3, with 90\% of galaxies fainter than 16. The magnitude distribution is shown in Figure~\ref{fig:magnitudes}, which also highlights the brightest and faintest detections.

\begin{figure*}[ht]
\centering{
\includegraphics[width=1\textwidth]{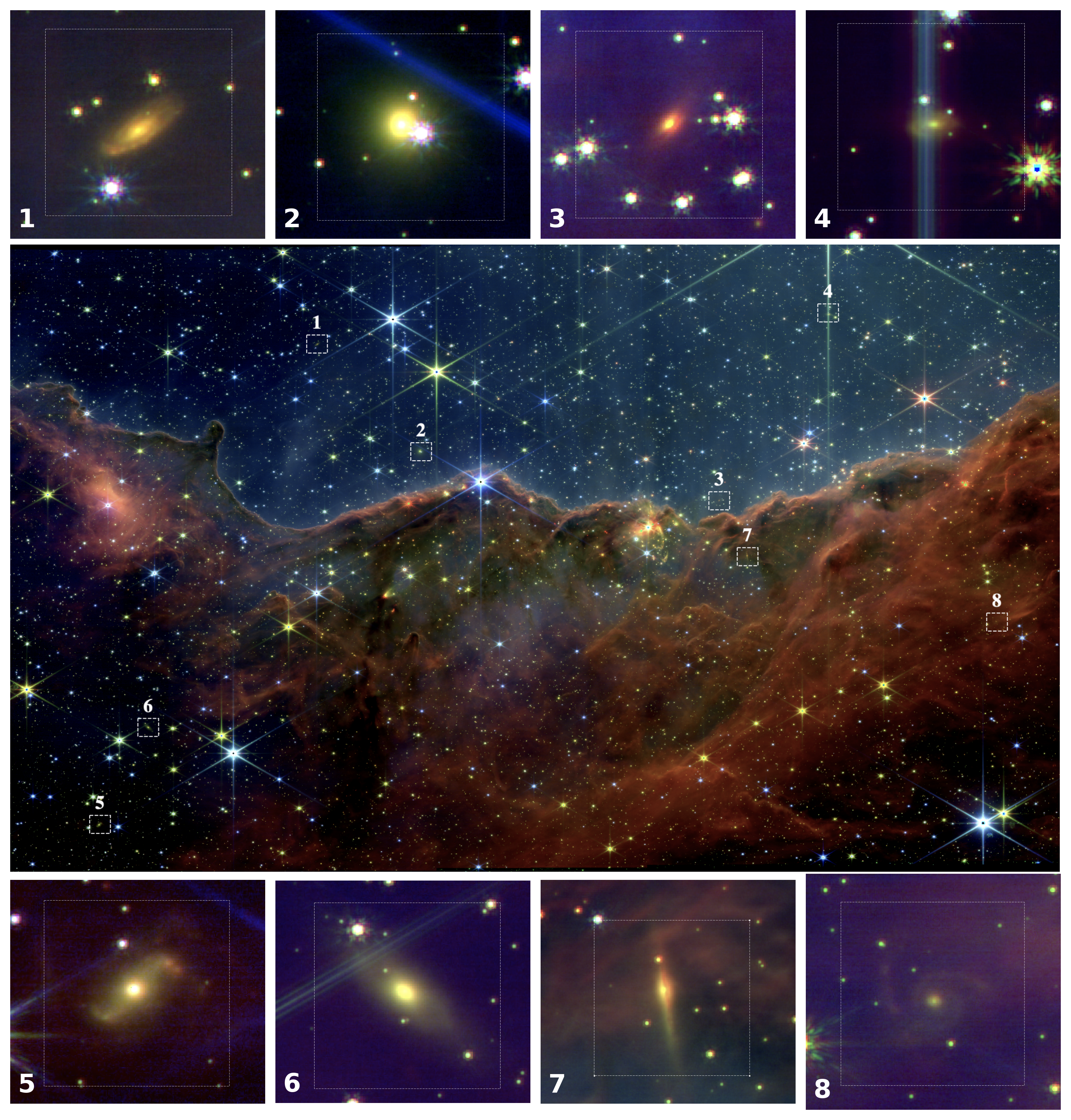}
\caption{Composite near-infrared view of the NGC~3324 region as observed by \jwst/NIRCam. The color image was constructed using the \texttt{F090W} (blue), \texttt{F200W} (green), and \texttt{F444W} (red) filters. The central panel displays the full field, revealing complex filamentary structures of ionized gas and embedded stars. The top and bottom rows show the eight brightest galaxies identified in the field, extracted from the same composite image and labeled 1 through 8 from top left to bottom right. Each stamp contains a white square of $5\arcsec \times 5\arcsec$ drawn for spatial reference. The figure is oriented in detector-native X/Y coordinates; North is rotated $102^\circ$ counterclockwise from vertical.}
}
\label{fig:rgb}
\end{figure*}

As a reference, the eight brightest galaxies are displayed in Figure~\ref{fig:rgb}, and their main properties are summarized below. The figure is oriented in detector X/Y pixel coordinates, corresponding to a rotation of 102$^\circ$ with respect to the standard astronomical orientation (north up, east left). Photometric measurements from both \texttt{AUTO} and \texttt{WIN} apertures are listed in Table~\ref{mags}.  

\textit{1. \jwst-ZOA~J159.2482$-$58.6502}: Bright face-on Sa spiral (\texttt{F444W} = 14.89). Two arms emerge clockwise from the nucleus, extending $\sim3\arcsec$, with bright knots along the northeastern arm. Morphology is clear in \texttt{F200W} and \texttt{F444W}, but only a faint nuclear component is visible in \texttt{F090W}.

\textit{2. \jwst-ZOA~J159.2300$-$58.6359}: Brightest and largest galaxy in the sample (\texttt{F444W} = 13.91; \texttt{ISOAREA} = 1967 pixel$^2$). Symmetric E0 elliptical with a diffuse halo extending $\sim4\arcsec$ in the redder bands. The nucleus is bright in \texttt{F444W} but absent in \texttt{F090W}.

\textit{3. \jwst-ZOA~J159.2341$-$58.6008}: Edge-on disk (\texttt{F444W} = 15.45; \texttt{ISOAREA} = 615 pixel$^2$), oriented west–east. A dust lane is seen in \texttt{F200W}, with a bright nucleus in \texttt{F444W}. Located $\sim5\arcsec$ east of the molecular cloud edge.

\textit{4. \jwst-ZOA~J159.2799$-$58.5932}: Sb spiral with north–south major axis $\sim4\arcsec$ and \texttt{F444W} = 14.02. The nucleus peaks in \texttt{F200W}. A nearby bright star introduces diffraction features, but the galaxy remains visible in all bands.

\textit{5. \jwst-ZOA~J159.1330$-$58.6624}: Sc spiral (\texttt{F444W} = 14.14), elongated NE–SW. A compact $\sim2\arcsec$ core dominates in \texttt{F200W} and \texttt{F444W}, while the extended disk is visible in all bands. Two bright knots terminate the SW arm, best seen in \texttt{F200W}.

\textit{6. \jwst-ZOA~J159.1565$-$58.6624}: Disk galaxy $\sim45\arcsec$ northeast of \jwst-ZOA~J159.1330$-$58.6624. The nucleus is detected in all bands, with $\mathrm{PA}\simeq68^\circ$ and axis ratio $b/a=0.73$. The disk spans $\sim3.6\arcsec$, aligned at $\mathrm{PA}\simeq60^\circ$, and is best defined in \texttt{F200W} and \texttt{F444W}.

\textit{7. \jwst-ZOA~J159.2235$-$58.5962}: Edge-on lenticular with a dust lane prominent in \texttt{F444W}. Despite lying in one of the densest regions of NGC~3324, it is the third brightest galaxy in the sample (\texttt{F444W} = 14.10). The disk is inclined by $\sim30^\circ$ relative to the detector $+X$ axis and appears sharply defined in \texttt{F200W}.

\textit{8. \jwst-ZOA~J159.2217$-$58.5662}: Face-on spiral $\sim3.5\arcsec$ across, showing three arms in a clockwise pattern. The nucleus peaks in \texttt{F200W}, while the \texttt{F444W} image emphasizes arm elongation and star-forming clumps.

\begin{table*}[ht]
\centering

\begin{tabular}{l l l c c c c c c}
\hline
Gx id  & \texttt{RA} & \texttt{Dec} &
$\mathrm{F090W}_{\mathrm{AUTO}}$ & $\mathrm{F090W}_{\mathrm{WIN}}$ &
$\mathrm{F200W}_{\mathrm{AUTO}}$ & $\mathrm{F200W}_{\mathrm{WIN}}$ &
$\mathrm{F444W}_{\mathrm{AUTO}}$ & $\mathrm{F444W}_{\mathrm{WIN}}$ \\
\hline
1 & 159.230006 & -58.635914 & 20.367 & 21.121 & 15.920 & 16.721 & 13.909 & 14.687 \\
2 & 159.279959 & -58.593268 & 21.711 & 22.441 & 17.530 & 18.543 & 14.024 & 14.788 \\
3 & 159.223594 & -58.596271 & 26.490 & 27.071 & 22.399 & 22.459 & 14.100 & 14.891 \\
4 & 159.133044 & -58.662467 & 21.391 & 22.161 & 16.116 & 16.872 & 14.149 & 14.918 \\
5 & 159.156587 & -58.659478 & 21.560 & 22.331 & 16.356 & 17.133 & 14.239 & 15.022 \\
6 & 159.248128 & -58.650276 & 24.504 & 25.164 & 18.397 & 19.421 & 14.891 & 15.667 \\
7 & 159.221713 & -58.566250 & \ldots & \ldots & 19.230 & 20.004 & 15.129 & 15.945 \\
8 & 159.234178 & -58.600848 & 24.815 & 25.661 & 18.585 & 19.347 & 15.449 & 16.239 \\
\hline
\end{tabular}
\caption{Multi-band photometric measurements for the eight brightest galaxies in the sample. The first column corresponds to the source identifiers shown in Figure~\ref{fig:rgb}. Coordinates are given in J2000. Magnitudes are reported in the {\tt AUTO} and {\tt WIN} apertures for the {\tt F090W}, {\tt F200W}, and {\tt F444W} bands. Missing entries indicate non-detections or measurements below the extraction threshold.}
\label{mags}
\end{table*}

\subsection{A Possible Compact Galaxy Association}

\begin{figure}[h]
    \centering
    \includegraphics[width=0.5\textwidth]{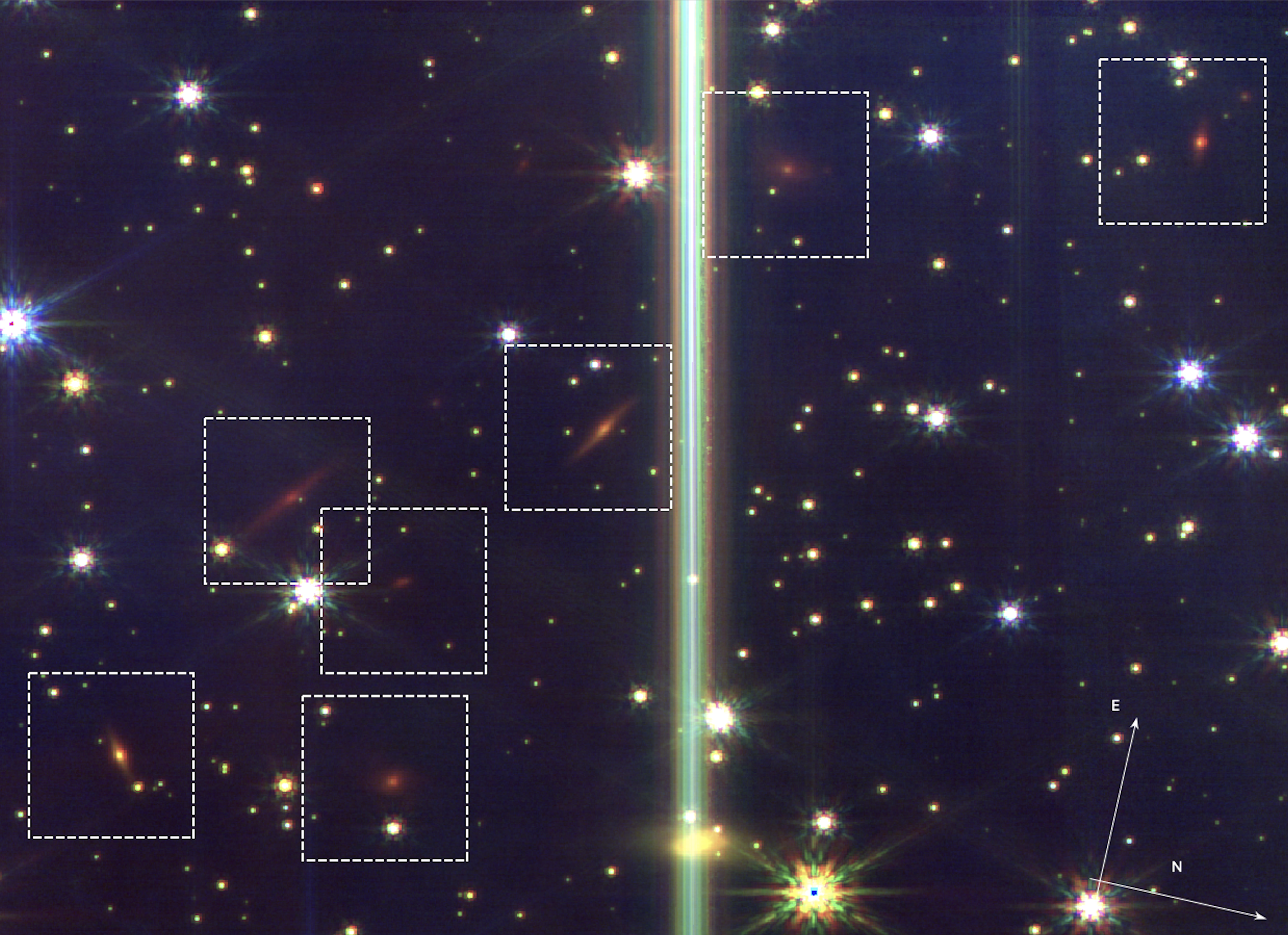}
    \caption{Composite RGB image of the eastern region of the field showing a compact overdensity of seven galaxies (dashed boxes) identified within a radius of $17''$ centered at (RA, Dec) = (159.2840$^\circ$, $-58.5948^\circ$). 
The image combines {\tt F200W}, {\tt F335M}, and {\tt F444W} filters. 
The bright vertical feature corresponds to a diffraction spike from a saturated star. 
The bright galaxy along a diffraction spike corresponds to Galaxy~4 (see Section~\ref{result}) and is not considered in the analysis. 
The figure is oriented in detector-native X/Y coordinates; North is rotated $102^\circ$ counterclockwise from vertical.
 }
    \label{fig:cluster}
\end{figure}

In the eastern region of the field, we identify a projected overdensity of seven galaxies within a rectangular area of $36.5'' \times 20.9''$, centered at (RA, Dec) = (159.2847$^\circ$, $-58.5939^\circ$), as illustrated in Figure~\ref{fig:cluster}. 
Each candidate is enclosed by a $4'' \times 4''$ box.

The galaxies exhibit comparable apparent brightness in the {\tt F444W} band, with six spanning $16.5 < m_{\mathrm{F444W}} < 16.9$~mag and one fainter source at $m_{\mathrm{F444W}} = 18.6$~mag. The median magnitude is $\langle m_{\mathrm{F444W}} \rangle = 16.7$ (IQR = 0.17~mag). Structural parameters show FWHM values between 3.4 and 9.9 pixels ($0.11''$–$0.31''$; assuming 0.032$''$~px$^{-1}$) and isophotal areas of 44–367 pixels. Ellipticities range from 0.27 to 0.79 (median = 0.46), consistent with inclined disk- or lenticular-like morphologies, although contamination from irregular or interacting systems cannot be excluded.  

As a preliminary reference, we explored order-of-magnitude redshift constraints. Adopting a characteristic half-light radius of 2.5–3.0~kpc for early-type systems \citep[e.g.,][]{Laurikainen2006,Laurikainen2011,Shen2003,vanDerWel2014} and a mean observed FWHM of $0.16''$, one infers an angular diameter distance of $\sim5$~kpc~arcsec$^{-1}$, corresponding to $z \sim 0.3$–0.4 in a flat $\Lambda$CDM cosmology ($H_0 = 70\ \mathrm{km\ s^{-1}\ Mpc^{-1}}$, $\Omega_m = 0.3$, $\Omega_\Lambda = 0.7$). However, this estimate remains highly uncertain due to the unknown extinction, intrinsic morphology, and possible projection effects.

Confirming the physical nature of this system requires spectroscopic follow-up to establish redshifts and assess whether the galaxies form a bound compact group or a chance superposition along low-extinction sightlines.

\subsection{Transnebular Sources}

One of the most compelling results of this study is the detection of galaxies behind the densest regions of the molecular cloud associated with NGC~3324, regions inaccessible to optical and even near-infrared surveys. We designate these objects as \emph{transnebular galaxies}.

Figure~\ref{fig:rgb} displays the full RGB composite of the region, where two such galaxies—sources 7 and 8 (bottom-right panels)—are highlighted. Both are located behind dense gas and dust columns, yet exhibit resolved morphologies.

Source 7 is classified as a lenticular galaxy. It presents a smooth, elongated, edge-on morphology with a prominent central bulge. The galaxy is bright (\texttt{F444W\_AUTO} = 15.13) and slightly reddened on its eastern side, but remains distinguishable against the molecular background.

Source 8, in contrast, is a face-on spiral galaxy with tightly wound arms encircling a bright nucleus. Despite lying behind one of the darkest regions of the cloud, both its core and spiral structure are confidently detected. The galaxy reaches a magnitude of \texttt{F444W\_AUTO} = 15.45.

The successful identification of these transnebular galaxies underscores the unprecedented extinction-penetrating capabilities of \jwst/NIRCam. It highlights its potential to reveal background structures even in the most obscured Galactic environments.

Unlike previous mid-infrared detections obtained with \textit{Spitzer}/IRAC or \textit{WISE}, which were limited by coarse angular resolution ($\sim2$--$6''$) and severe confusion in crowded Galactic regions, the \textit{JWST}/NIRCam data provide a tenfold improvement in spatial resolution and nearly two orders of magnitude deeper sensitivity. This combination enables the direct morphological identification of background galaxies through the densest molecular filaments of NGC~3324---regions entirely inaccessible to earlier facilities. The precedent established here is thus observational rather than instrumental: it demonstrates, for the first time, that extragalactic sources can be resolved and morphologically classified through opaque Galactic clouds, validating the feasibility of systematic \textit{JWST} searches across the Zone of Avoidance.

\section{Results and Discussion} \label{discusion}

This study leverages the unprecedented sensitivity and spatial resolution of \jwst\  to assess the detectability of background galaxies through the heavily obscured star-forming region NGC~3324. Using wide-band NIRCam imaging, we conducted a comprehensive photometric and morphometric analysis of over 18,000 sources. Source extraction and classification, performed via \texttt{SExtractor}, enabled the identification of 102 galaxies, many of which remain beyond the reach of previous-generation instruments. This constitutes the first systematic demonstration that \jwst\ can penetrate the ZoA, enabling extragalactic exploration through highly extinct Galactic environments.

Despite being originally conceived for less complex observational regimes, \texttt{SExtractor} proved to be a robust tool under the extreme background conditions characteristic of the Galactic plane. By tuning rarely explored parameters, we optimized source detection across structured and diffuse emission fields. Critical adjustments included setting \texttt{BACK\_PEARSON} to 3.5 to suppress correlated interpolation artifacts, increasing \texttt{BACKPHOTO\_THICK} to 12 pixels to stabilize local background estimates, and applying \texttt{BACK\_FILTTHRESH} = 3 for selective filtering. These were combined with a background mesh of 32 pixels and a filter size of 3. This configuration enabled the detection of both faint sub-arcsecond galaxies and extended low-surface-brightness sources, while spurious detections were mitigated through stringent flux and area thresholds.

To validate the photometric reliability of the extracted catalog, we compared \texttt{SExtractor}-derived fluxes against PSF photometry computed with \texttt{DAOPHOT}. Across the NIRCam bands, the agreement was excellent, with typical offsets below 0.02~mag and standard deviations of order 0.01~mag. These results confirm the method's reliability under high-background, spatially variable conditions.

Using these high-fidelity measurements, we established an empirical morphological classification scheme based on the joint distribution of \texttt{FWHM} and \texttt{SNR\_WIN}. By applying non-parametric criteria derived from boxplot statistics ($\texttt{FWHM} < 3.124$ and $\texttt{SNR\_WIN} > 29.257$), we partitioned the parameter space into four morpho-photometric sectors. This segmentation enables the isolation of physically distinct populations and supports subsequent population synthesis analyses.

The detected galaxies span a broad range of magnitudes and structural profiles. The brightest source reaches $\texttt{F444W\_AUTO} \sim 13.9$~mag, while the faintest lies near $\texttt{F444W\_AUTO} \sim 19.3$~mag. Notably, 90\% of the sample exhibits magnitudes fainter than $m_{\mathrm{F444W}} \sim 17$, consistent with a distant and extinction-attenuated population.

As expected, the spatial distribution of detected galaxies is strongly modulated by the internal extinction profile of the NGC~3324 molecular cloud. A total of 85 galaxies are located within the central cavity, with the remainder distributed across two peripheral low-extinction regions in the northwestern and southwestern corners of the field.

We also report the discovery of a compact galaxy association in the eastern field, exhibiting a significant overdensity relative to the expected background. From the observed angular sizes and inferred morphologies of its members—primarily lenticular—we estimate a redshift of $z \sim 0.4$. This suggests the presence of a gravitationally bound, intermediate-redshift group projected behind the Galactic plane. If confirmed via spectroscopic follow-up, this system may represent the most distant galaxy cluster detected within the ZoA.

Among the most notable results is the identification of \textit{transnebular galaxies}—sources detected through the densest regions of the molecular cloud. Their detection establishes a precedent for penetrating even the most opaque Galactic structures. Morphological analyses confirm that both spiral and lenticular galaxies can be resolved under these conditions, underscoring the transformative potential of \jwst\ for accessing the obscured extragalactic sky.

\begin{figure}[h]
    \centering
    \includegraphics[width=0.48\textwidth]{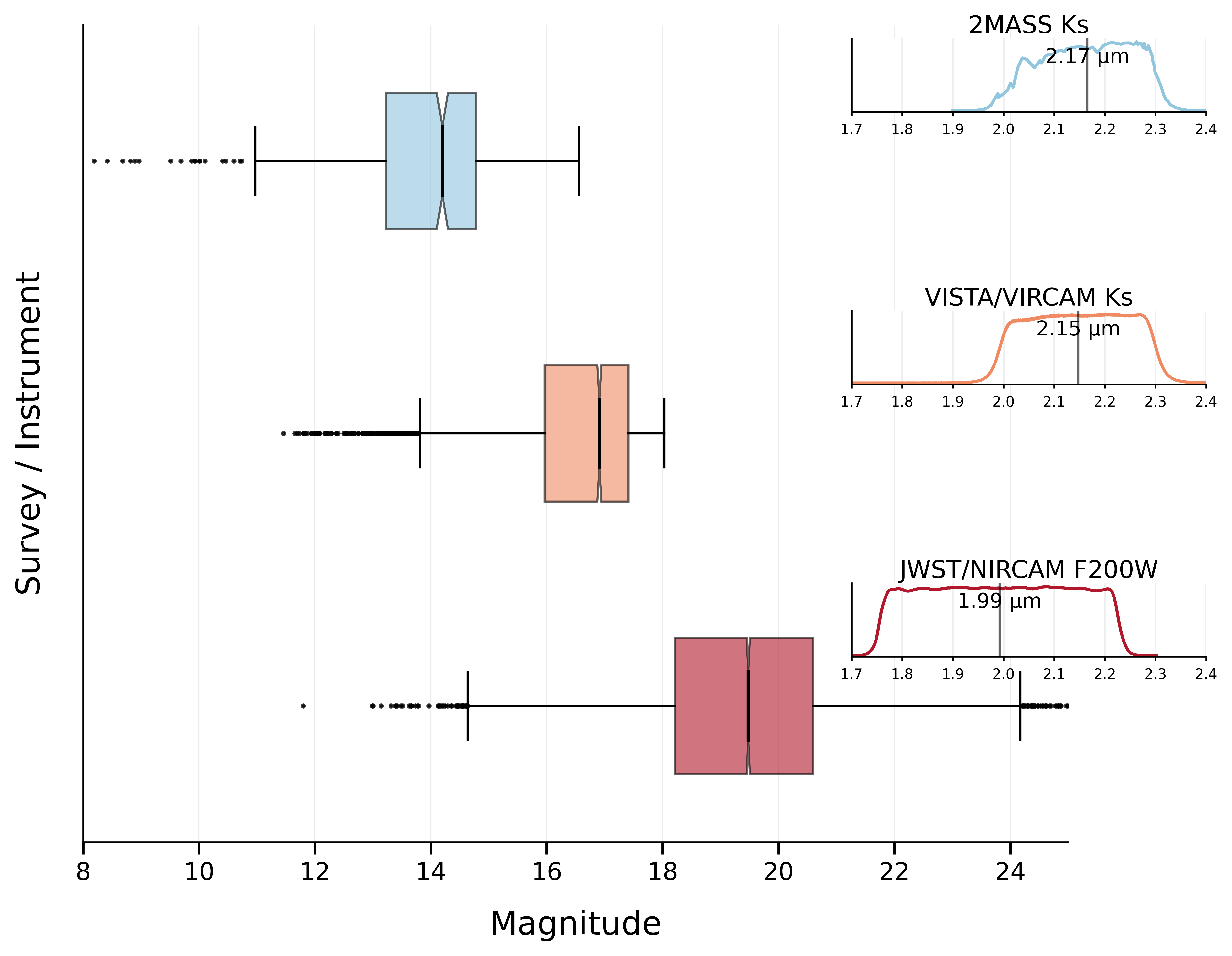}
    \caption{Magnitude distributions for \jwst/NIRCam~F200W, VISTA/VIRCAM~$K_s$, and 2MASS~$K_s$ sources in the NGC~3324 field. Boxes show interquartile ranges with notched medians, and insets display the normalized transmission curves and effective wavelengths of each filter.
}
    \label{fig:comparacion01}
\end{figure}

\begin{figure}[h]
    \centering
    \includegraphics[width=0.48\textwidth]{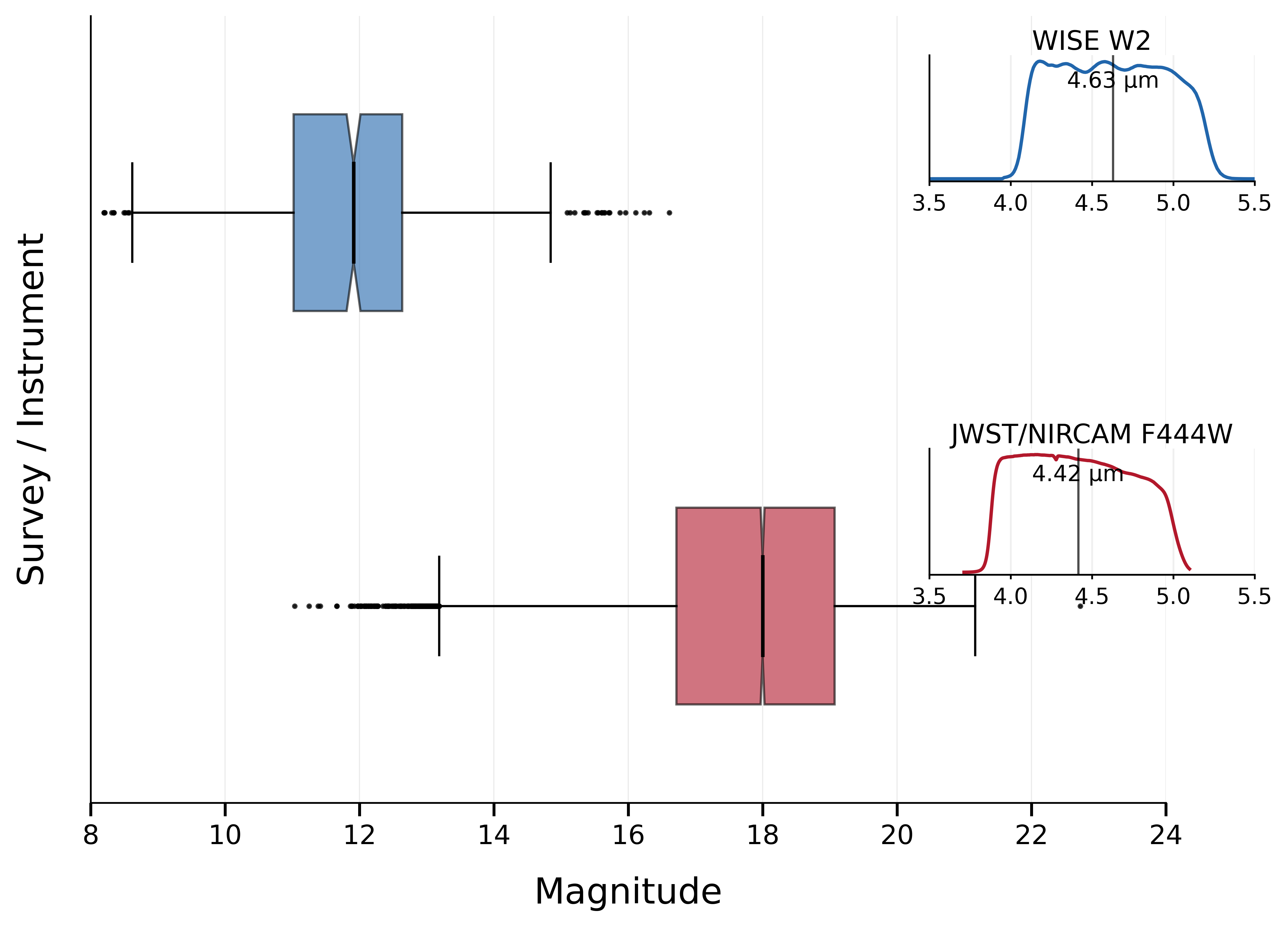}
\caption{Same as Fig.~\ref{fig:comparacion01}, but for \jwst/NIRCam~F444W and WISE/W2 sources in the NGC~3324 field. }
    \label{fig:comparacion02}
\end{figure}

To quantify the incremental gain achieved by \textit{\jwst} relative to previous infrared facilities, we compared our NIRCam detections against sources from 2MASS, VISTA/VIRCAM (VVVX), and WISE \citep{Wright2010} within the same field. Figure~\ref{fig:comparacion01} and Figure~\ref{fig:comparacion02} summarizes these comparisons, showing the magnitude distributions together with the corresponding filter transmission curves.

The near--infrared comparison between the $K_s$--band surveys and NIRCam/F200W reveals a consistent and substantial depth extension. Median magnitudes progress from $14.2$\,mag in 2MASS to $16.9$\,mag in VISTA/VIRCAM, reaching $19.5$\,mag in \jwst/F200W. The interquartile ranges (1.55, 1.45, and 2.39\,mag, respectively) indicate the increasing dynamical span of the detected populations. At the 98th percentile, the F200W distribution extends 4.8\,mag deeper than VISTA and 6.8\,mag deeper than 2MASS. The uniformity and stability of the \jwst point spread function further allow the detection of diffuse, low--surface--brightness structures that remain unresolved in ground--based images.

A similar result arises in the mid--infrared regime when comparing NIRCAM/F444W with WISE/W2 (Figure~\ref{fig:comparacion02}). The median magnitude of the F444W sample is $18.0$\,mag versus $15.3$\,mag for WISE, representing an improvement of $+2.7$\,mag. At the bright end, both datasets show consistent zero points, but at fainter levels the \jwst\ distribution broadens markedly, with detections extending $\sim1.5$\,mag beyond the WISE completeness limit. The narrower IQR of the WISE sample ($1.62$\,mag) compared to NIRCam ($2.35$\,mag) reflects the strong confusion and photometric saturation affecting WISE in crowded Galactic regions.

In the specific case of the NGC~3324 field, the VVV~NIRGC catalogue reports no extragalactic sources despite full coverage. The absence of detections is therefore attributable to the intrinsic limitations of ground--based infrared imaging under severe crowding and extinction, rather than to survey incompleteness. For reference, an identical composition to Figure~\ref{fig:rgb} constructed from VVVX data is presented in Appendix~\ref{fig:vvvx}. 

Together, these comparisons establish a continuous photometric bridge between \jwst\ and the main legacy infrared datasets. The measured offsets and completeness gains provide an empirical basis to cross--calibrate \jwst\ fluxes with historical all--sky surveys, ensuring both backward compatibility and a quantifiable extension of the detection frontier into the low--surface--brightness, high--extinction regime. 

In this framework, the forthcoming \textit{Nancy Grace Roman Space Telescope} \citep{roman} will play a complementary role. Its wide--field near--infrared imaging, though limited to wavelengths shorter than $\sim2\,\mu$m, will deliver homogeneous, deep coverage over thousands of square degrees. When combined with \jwst’s sub--arcsecond resolution and spectral reach, Roman’s surveys will enable a dual--scale mapping of obscured regions: \textit{\jwst} resolving the faintest and most compact extragalactic components, and \textit{Roman} establishing the statistical and cosmographic continuity across the Galactic plane. The synergy between both facilities will thus define the next observational standard for extragalactic reconstruction through the Zone of Avoidance.

 \begin{acknowledgements}
J.L.N-C. is grateful to the Universidad de La Serena for providing the academic environment and support that allowed the development of this research. This work was conducted without specific financial support, a fact mentioned here simply for clarity. L.D.B., M.V.A. and C.V. thank the support of the Consejo de Investigaciones Cient\'ificas y T\'ecnicas (CONICET) and Secretar\'ia de Ciencia y T\'ecnica de la Universidad Nacional de C\'ordoba (SeCyT). C.N.A.W is supported by contract \jwst/NIRCam  NAS5-02015 to the University of Arizona. F.M.C. thanks the support of ANID BECAS/DOCTORADO NACIONAL 21110001. D.M. gratefully acknowledges support by the ANID BASAL projects ACE210002 and FB210003 and by Fondecyt Project No. 1220724. M.S. acknowledges support by ANID’s Fondecyt Regular Project \#1251401. I.V Daza-Perilla acknowledges funding by NASA under the CRESST II program.
This work is based on observations made with the NASA/ESA/CSA James Webb Space Telescope. The data were obtained from the Mikulski Archive for Space Telescopes at the Space Telescope Science Institute, which is operated by the Association of Universities for Research in Astronomy, Inc., under NASA contract NAS 5-03127 for \jwst. These observations are associated with program \# 2731. The authors gratefully acknowledge data from the ESO Public Survey program IDs 179.B-2002 and 198.B-2004 taken with the VISTA telescope, and products from the Cambridge Astronomical Survey Unit (CASU).  This research has made use of the NASA/IPAC Infrared Science Archive, which is funded by the National Aeronautics and Space Administration and operated by the California Institute of Technology

\end{acknowledgements}
\bibliographystyle{aa} 
\bibliography{refs} 

\begin{appendix}
\onecolumn

\section{Mosaic of Detected Galaxies in the F444W Band}
\begin{figure}[ht]
\centering
\includegraphics[width=\textwidth]{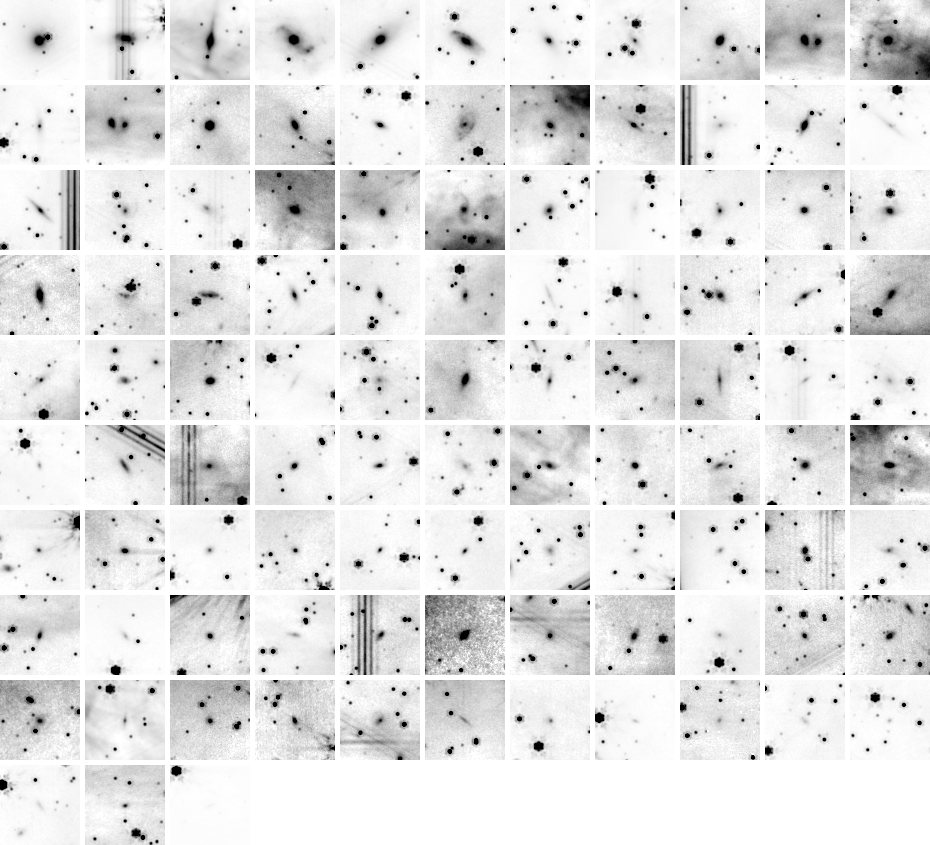}
\caption{
Mosaic of all 102 galaxies detected in the field from the {\tt F444W} reference image (see Section~\ref{catalogs}). 
Each stamp measures $3'' \times 3''$. 
The galaxies are arranged in order of decreasing brightness, from the brightest object in the upper-left corner to the faintest in the lower-right corner. 
The figure highlights the morphological diversity of the sample, including edge-on disks, spheroidal systems, and irregular galaxies.
}
\label{fig:mosaic}
\end{figure}

\newpage

\section{Comparison with Ground-based VVVX Imaging}
\begin{figure}[ht]
\centering
\includegraphics[width=\textwidth]{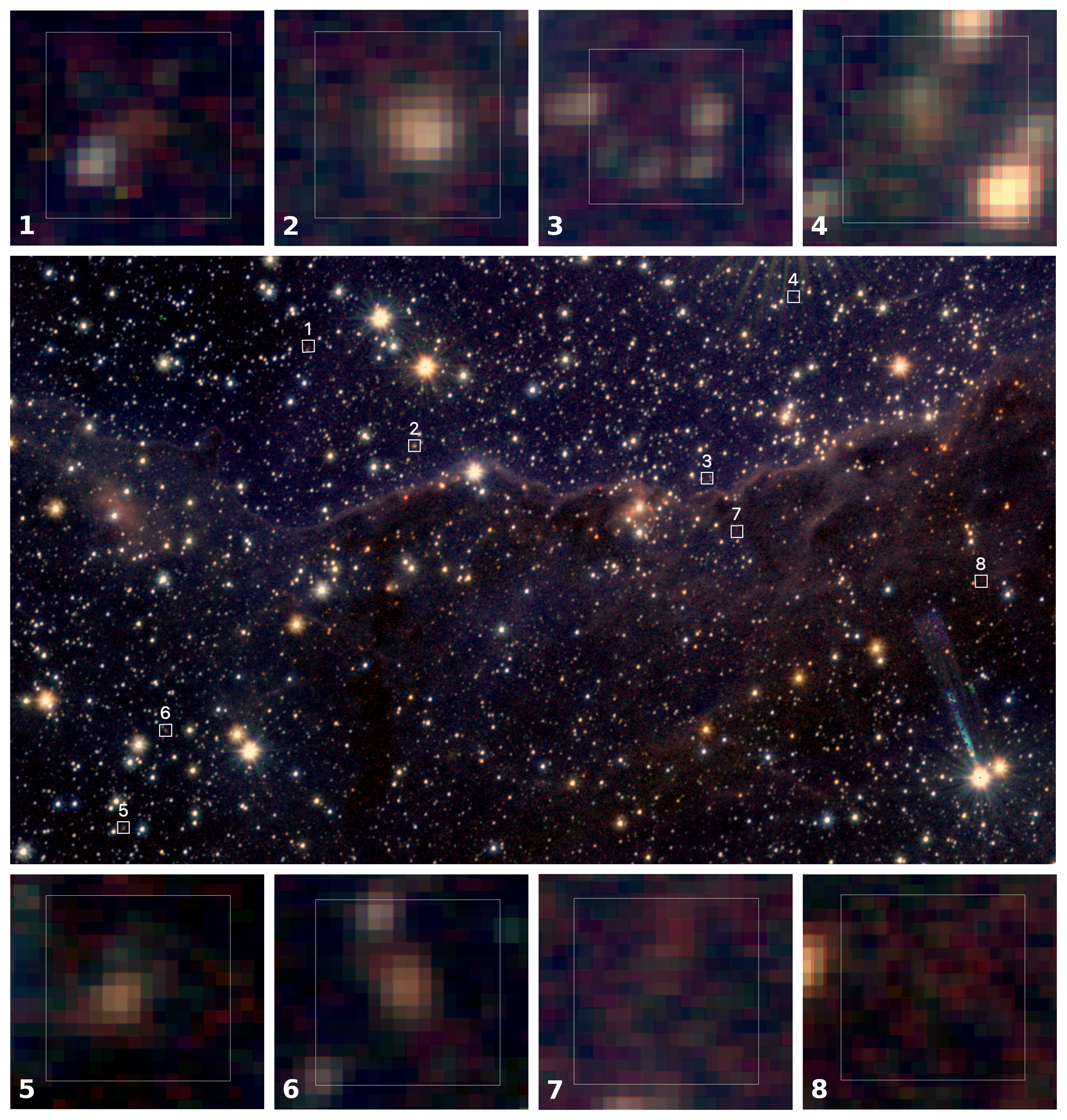}
\caption{
Composite near-infrared view of the NGC~3324 region as observed by the VVVX survey. 
The field corresponds to tile {\tt e1084}, observed on March~19,~2018, and fully covered in the $J$, $H$, and $K_s$ bands. 
Despite being located within the nominal depth of the survey, no galaxies were included in the VVV–NIRGC catalogs. 
The central panel displays the same field shown in Fig.~\ref{fig:rgb}, preserving its orientation to enable a direct comparison with the \jwst/NIRCam image. 
The top and bottom rows show the eight brightest galaxies (1–8) identified in the \jwst image, placed at their corresponding sky positions within the VVVX frame. 
Each stamp contains a white $5'' \times 5''$ box for spatial reference. 
North is rotated $102^\circ$ counterclockwise from vertical, matching the orientation adopted in Fig.~\ref{fig:rgb}. 
This comparison highlights the observational limitations of ground-based near-infrared imaging in regions of high extinction and stellar crowding, and the substantial gain in sensitivity and resolution achieved with \jwst.
}
\label{fig:vvvx}
\end{figure}

\twocolumn
\end{appendix}

\end{document}